\documentclass[draft]{agujournal2019}
\usepackage{url} 
\usepackage{lineno}
\usepackage{siunitx}
\usepackage{soul}
\draftfalse

\journalname{JGR: Oceans}

\begin{document}

%
%

\title{Frequency dependence of near-surface oceanic kinetic energy from drifter observations and global high-resolution models}

%
%




\authors{Brian K. Arbic\affil{1,*}, Shane Elipot\affil{2,*}, Jonathan M. Brasch\affil{3}, Dimitris Menemenlis\affil{4},  Aur\'elien L. Ponte\affil{5}, Jay F. Shriver\affil{6}, Xiaolong Yu\affil{5,+}, Edward D. Zaron\affil{7}, Matthew H. Alford\affil{8}, Maarten C. Buijsman\affil{9}, Ryan Abernathey\affil{10}, Daniel Garcia\affil{3}, Lingxiao Guan\affil{3}, Paige E. Martin\affil{10,11}, Arin D. Nelson\affil{1,12} \vspace{0.4in} {*These authors contributed equally to this study.\\+Now at:  School of Marine Sciences, Sun Yat-sen University, Zhuhai, China}
}


\affiliation{1}{Department of Earth and Environmental Sciences, University of Michigan, Ann Arbor, MI, USA}
\affiliation{2}{Rosenstiel School of Marine and Atmospheric Science, University of Miami, Miami, FL, USA}
\affiliation{3}{Department of Electrical Engineering and Computer Science, University of Michigan, Ann Arbor, MI, USA}
\affiliation{4}{Jet Propulsion Laboratory, California Institute of Technology, Pasadena, CA, USA}
\affiliation{5}{Ifremer, Université de Brest, CNRS, IRD, Laboratoire d'Océanographie Physique et Spatiale, IUEM, Brest, France}
\affiliation{6}{Naval Research Laboratory, Stennis Space Center, Mississippi,USA}
\affiliation{7}{College of Earth, Ocean, and Atmospheric Sciences, Oregon State University, Corvallis, OR, USA}
\affiliation{8}{Scripps Institution of Oceanography, University of California San Diego, La Jolla, California, USA}
\affiliation{9}{School of Ocean Science and Engineering, University of Southern Mississippi, Stennis Space Center, Mississippi}
\affiliation{10}{Lamont-Doherty Earth Observatory, Columbia University, New York City, NY, USA}
\affiliation{11}{Department of Physics, University of Michigan, Ann Arbor, MI, USA}
\affiliation{12}{Graduate School of Oceanography, University of Rhode Island, Kingston, RI, USA}

\vspace{0.3in}




\correspondingauthor{Brian K. Arbic $\&$ Shane Elipot}{arbic@umich.edu $\&$ selipot@miami.edu}




\begin{keypoints}
\item We examine frequency content of ocean kinetic energy (KE) at the sea surface (0 m) and 15 m depth with global drifter data and two models. 
\item Modeled near-inertial and tidal kinetic energy values are sensitive to wind forcing frequency and parameterized damping, respectively.  
\item Models capture reasonably well the spatial- and frequency-dependence of the observed ratio of 0 m KE to (0 m + 15 m) KE.

\end{keypoints}




%
%

%
%


\begin{abstract}

The geographical variability, frequency content, and vertical structure of near-surface oceanic kinetic energy (KE) are important for air-sea interaction, marine ecosystems, operational oceanography, pollutant tracking, and interpreting remotely sensed velocity measurements.  Here, KE in high-resolution global simulations (HYbrid Coordinate Ocean Model; HYCOM, and Massachusetts Institute of Technology general circulation model; MITgcm), at the sea surface (0 m) and 15 m, are respectively compared with KE from undrogued and drogued surface drifters.  Global maps and zonal averages are computed for low-frequency ($<$ 0.5 cpd), near-inertial, diurnal, and semi-diurnal bands.  Both models exhibit low-frequency equatorial KE that is low relative to drifter values.  HYCOM near-inertial KE is higher than in MITgcm, and closer to drifter values, probably due to more frequently updated atmospheric forcing.  HYCOM semi-diurnal KE is lower than in MITgcm, and closer to drifter values, likely due to inclusion of a parameterized topographic internal wave drag.  A concurrent tidal harmonic analysis in the diurnal band demonstrates that much of the diurnal flow is non-tidal.  We compute a simple proxy of near-surface vertical structure, the ratio of 0 m KE to 0 m KE plus 15 m KE in model outputs, and undrogued KE to undrogued KE plus drogued KE in drifter observations.  Over most latitudes and frequency bands, model ratios track the drifter ratios to within error bars.  Values of this ratio demonstrate significant vertical structure in all frequency bands except the semidiurnal band.  Latitudinal dependence in the ratio is greatest in diurnal and low-frequency bands.  
   
\end{abstract}

\section*{Plain Language Summary}

It is important to map and understand ocean surface currents because they affect climate and marine ecosystems.  Recent advances in global ocean models include the addition of astronomical tidal forcing alongside atmospheric forcing, and the usage of more powerful computers that can resolve finer features.  Here we evaluate ocean surface currents in high-resolution simulations of two different ocean models through comparison with observations from surface drifting buoys.  We examine near-inertial motions, forced by fast-changing winds, semidiurnal tides, forced by the astronomical tidal potential, diurnal motions, arising from tidal and other sources, and low-frequency currents and eddies, forced by atmospheric fields.  Global patterns in the models and drifters are broadly consistent.  The two models differ in their degree of proximity to drifter measurements in the near-inertial band, most likely due to different update intervals for atmospheric forcing, and in the semidiurnal band, most likely due to different damping schemes.  A simple proxy for vertical structure of the currents, measured by differences in drifter flows at the surface vs. 15 meters depth, is tracked reasonably well by the models.  Discrepancies between models and observations motivate future improvements in the models.

%
%

%


%
%
%
%

\section{Introduction}

This paper presents a comprehensive comparison of the geographical distribution and frequency content of near-surface kinetic energy (KE) from two different global high resolution models with observations from NOAA's Global Drifter Program (GDP).  We divide KE into different frequency bands of interest via spectral analysis.  We compare model results at the sea surface with results from undrogued drifters, and model results from 15 m depth with results from drogued drifters.  Our ability to examine a wide range of frequencies is enabled by the employment of astronomical tidal forcing alongside atmospheric forcing in both models. We compute a proxy ratio of vertical structure, the ratio of KE at the surface divided by the sum of surface KE and 15 m KE, and compare this ratio in the models vs drifters.  The proxy ratio that we use is an incomplete measure of vertical structure.  It is emphasized here because it can be computed on a global-scale from drifter observations, thus offering a novel global-scale examination of the ability of models to simulate aspects of vertical KE structure in the upper ocean.  In the near future, the global near-surface ocean velocity field may be measured by remote sensing missions, such as the potential ``Odysea” mission.  Until such missions are launched, models such as those used here, and drifter observations, will be the best way to map the global ocean surface velocity field over a wide range of frequency bands.  Even after such missions are launched, drifter observations will serve as an important ``ground truth” dataset, and models will continue to be important interpretive tools.  
This manuscript aims at taking a wide-ranging, holistic descriptive approach.  There are many sources of difference between the two models used, and between the models and drifter data.  In addition, there are many important sensitivities and processes that remain to be explored, far more than can be examined in one paper.  We will note the multiple sources of difference and unexplored sensitivities and processes as they arise throughout the manuscript.  A major aim of this manuscript is to motivate future studies on particular aspects of the model-observation and model-model discrepancies that we document here.  We believe that a paper focused on a broad-brush analysis is well-suited for highlighting potential areas for process studies and model improvements. 

Oceanic surface currents are relevant for a range of multi-disciplinary scientific topics and operational applications \cite{ElipotWenegrat2021}.  For instance, surface currents are major actors in two crucial components of the climate system, the air-sea transition zone \cite{Cronin2019} and marine ecosystems \cite{Levy2018}.  Maps of surface currents are useful for understanding ocean dynamics, assessing operational ocean modeling, enabling search-and-rescue missions, and tracking oil spills, plastics, and other marine pollutants.  Many of these applications, including predicting Lagrangian trajectories of water parcels in the near-surface ocean, understanding air-sea transfer, and others, require knowledge not only of currents but also their vertical structure near the oceanic surface \cite{Beron2019building}.  

Despite the importance of near-surface currents, not enough is known about their space-time variability.  Mapping near-surface oceanic currents on a global scale is a formidable task.  Quantifying high-frequency motions such as near-inertial flows \cite<e.g.,>{Pollard1970,Alford2003a,Alford2003b,Furuichi2008,chaigneau2008global,Simmons2012}, semidiurnal tides, diurnal tides and other diurnal motions, and the internal gravity wave continuum \cite{GarrettMunk1975} requires high-frequency sampling (e.g., at approximately hourly intervals).  Time series from moored current meters can be used to separate high-frequency motions from lower-frequency motions including Ekman flows, mesoscale eddies, and the oceanic general circulation.  Another advantage of mooring lines is the delivery of observations below the ocean surface.  However, moored data are expensive to deploy and are available only at a relatively small number of geographical locations \cite<see, for instance, Figure 1 of>{Luecke2020}.  Global-scale surface currents can be computed from satellite altimeter measurements of sea-surface height (SSH), providing that tides are removed and the geostrophic assumption is applied to the tide-corrected SSH fields.  However, altimetry cannot detect near-inertial or Ekman flows, due to the negligible SSH signal in these motions.  Altimetry measurements are also infrequent in time.  The repeat time of the TOPEX/Jason series, for instance, is about 10 days.  The geostrophic velocities computed from altimetry leave out high-frequency contributions to the velocity and hence KE fields. 

The drifting buoys, or drifters, of the GDP \cite{LumpkinOzgokmenCenturioni2017} yield a global dataset of near-surface ocean velocity in situ estimates. Drifters have been used in many previous studies of low-frequency oceanic flows, including the mean dynamic topography \cite{Maximenko2009}, Ekman flows \cite{Elipot2009}, and the eddying general circulation \cite{Thoppil2011}.  Recently, drifters have also been used to quantify high-frequency motions such as tides \cite{PoulainCenturioni2015,Kodaira2015,ZaronElipot2021}.  \citeA{Elipot2016} derived an hourly drifter product and demonstrated that it resolves motions at a wide range of frequencies.  

Model-drifter comparison over a wide range of frequency bands is now possible, due to the emergence of a new class of high-resolution global ocean models that simultaneously include astronomical tidal forcing and atmospheric forcing \cite<e.g.,>{Arbic2010,Arbic2012,Arbic2018,Arbic2022, Buijsman2020, Muller2012, Waterhouse2014, Rocha2016a, Rocha2016b}.  In these models, internal tides and mesoscale eddies co-exist and interact \cite{Shriver2014,Buijsman2017,Nelson2019}.  As shown first in \citeA{Muller2015} and later in other studies \cite<e.g.,>{Rocha2016a, Savage2017a, Savage2017b, Qiu2018,Torres2018,Luecke2020}, such models are beginning to partially resolve the internal gravity wave (IGW) continuum \cite<Garrett-Munk spectrum;>{GarrettMunk1975}.  \citeA{Yu2019} compared KE, over various low- and high-frequency bands, from the hourly drifter dataset and output from a high-resolution Massachusetts Institute of Technology general circulation model \cite<MITgcm;>{Marshall1997} simulation, forced by both the astronomical tidal potential and atmospheric fields.  The MITgcm simulation employed in \citeA{Yu2019}, designated as LLC4320, is further used here in a three-way comparison with drifter observations and a global internal tide and gravity wave simulation of the HYbrid Coordinate Ocean Model \cite<HYCOM;>{Chassignet2009}, the backbone operational model of the United States Navy.    

Global IGW model simulations, especially MITgcm LLC4320 and from HYCOM, have been widely used by the community, to plan for field campaigns \cite<e.g.,>{Wang2018}, understand interactions between motions at different length and time scales \cite<e.g.,>{Pan2020}, and provide boundary forcing for higher-resolution regional models \cite{Nelson2020}.  HYCOM and MITgcm LLC4320 have been used to quantify the relative contributions of low- and high-frequency motions to SSH and KE as a function of geographical location \cite<e.g.,>{Richman2012, Savage2017a,Savage2017b, Rocha2016a, Rocha2016b, Qiu2018,Torres2018}.  The energetics of different classes of oceanic motions are of interest in their own right \cite{ferrariwunsch2009}.  Global IGW models offer the potential for examining energy exchanges between different classes of oceanic motions, as has been seen in observations \cite{leboyeralford2021}.  Quantitative mapping of low- and high-frequency motions is important for satellite missions including the Surface Water Ocean Topography (SWOT) mission \cite{Morrow2019}, planned for launch in 2022, which will measure SSH at high resolution in two-dimensional swaths.  Remote sensing missions focused on measuring near-surface ocean velocities, such as the existing airborne Sub-Mesoscale Ocean Dynamics Experiment (S-MODE) mission \cite{Rodriguez2020}, and proposed velocity-measuring satellite missions \cite{Ardhuin2019,Rodriguez2020}, now under the umbrella name ``Odysea", will benefit from quantification of high- and low-frequency KE as well.  Because HYCOM and MITgcm LLC4320 are widely used, it is important to compare these models to observations.  A summary of model comparisons of HYCOM and MITgcm LLC4320 tidal simulations with mooring and altimeter observations is provided in \citeA{Arbic2018} and \citeA{Arbic2022}.  MITgcm LLC4320 has also been compared to along-track Acoustic Doppler Current Profiler (ADCP) data \cite{Rocha2016a,Chereskin2019}.

As demonstrated by \citeA{Yu2019}, drifters are an excellent dataset for revealing strengths and weaknesses of numerical ocean simulations.  For instance, \citeA{Yu2019} showed that LLC4320 semidiurnal tidal KE was too strong, and near-inertial KE too weak, relative to drifter observations.  Here, we build upon their study through intercomparison of HYCOM, MITgcm LLC4320, and drifters.  We employ an udpated analysis of the drifters, that yields fewer spatial gaps than the analysis employed in \citeA{Yu2019}, especially near the equator.  The HYCOM and MITgcm LLC4320 simulations differ in several respects, and we anticipate that they will perform differently in comparisons to drifters.  We especially anticipate differences in the near-inertial bands, due to more frequent updates of the wind fields in HYCOM (3 hours) relative to MITgcm LLC4320 (6 hours), and in the semidiurnal tidal band, due to the lack of a parameterized topographic wave drag in MITgcm LLC4320.  In HYCOM simulations, a parameterized topographic internal wave drag is included in order to roughly account for the damping of tidal motions due to breaking small-scale internal tides that are unresolved in global models \cite{Arbic2010,Arbic2018,Ansong2015,Buijsman2016,Buijsman2020,Arbic2022}.  \citeA{Ansong2015} demonstrated that the SSH signature of internal tides in HYCOM is closer to altimetry observations when the HYCOM simulations contain a wave drag than when they do not.  Here we examine the impact of including wave drag (as in HYCOM) vs. excluding it (as in LLC4320) on near-surface semidiurnal KE.  The diurnal band is also of special interest.  Here, we complement the frequency spectral analysis of the diurnal band in the models with a tidal harmonic analysis.  Differences between maps computed from frequency spectra versus tidal harmonic analysis can reveal whether the diurnal band also has significant non-tidal sources in the models.

An examination of the vertical structure of near-surface currents will aid our understanding of the air-sea exchange of heat, momentum, and gases, and of the dispersal of pollutants and biologically important tracers \cite{ElipotWenegrat2021}.  The vertical structure of velocity has important implications for the ongoing S-MODE airborne mission and proposed satellite missions focusing on surface ocean velocity measurements \cite{Ardhuin2019,Rodriguez2020}.  Such missions will need information on the frequency dependence of vertical structure in order to interpret the implications of surface current measurements for subsurface oceanic conditions.  

Here we examine a vertical structure proxy ratio, based upon the relative strengths of KE at the sea surface and at 15 m depth, the depths accessible to undrogued and drogued drifters.  To our knowledge, a global comparison of the 0 and 15 m KE, for both high- and low-frequency motions, has not been done before.  As in \citeA{Yu2019}, we use both undrogued and drogued drifters which provide estimates of oceanic velocity at 0 m (sea surface) and 15 m, respectively.  Accordingly, we compare model results at 0 m to undrogued drifter results, and model results at 15 m to drogued drifter results.  Error bars on the vertical structure proxy ratio in drifter observations are large, especially for near-inertial and tidal motions, and windage (erroneous slips of water past drifters) are a known problem in undrogued drifter observations (Section 2.4).  Nevertheless, the model ratios, which suffer from completely different biases and errors, follow the drifter ratios over most latitudes and most frequency bands.

The paper is organized as follows.  Section 2 describes the HYCOM simulation, the MITgcm LLC4320 simulation, and the drifter observations. This section also discusses the vertical momentum mixing schemes impacting modeled vertical structure in the upper ocean, and describes the data analysis methods.  The main results of the paper, consisting of rotary frequency spectra and maps and zonal averages of KE over specific frequency bands, are provided in Section 3.  A discussion of the results is put forth in Section 4, while Section 5 provides a summary and concluding remarks.
      
\section{Data and Methods}

We begin this section with a brief discussion of the limitations inherent in running and analyzing large global simulations.  The simulations used in this study are computationally expensive and finding multiple years of output to analyze is not yet possible.  Further, the individual years examined in this study are different for HYCOM and MITgcm LLC4320.  The latter simulation was a ``one-off", run for just over a year, and is, to the best of our knowledge, the most computationally expensive ocean numerical simulation ever undertaken.  The HYCOM simulations, on the other hand, comprise the hydrodynamical backbone of US Navy ocean forecasting, and as such are constantly being improved and re-run.  To find HYCOM simulations run during the same year as the MITgcm LLC4320 simulation, one would have to return to older HYCOM simulations, which have known drawbacks that have since been fixed.  The centers that run these simulations do not coordinate with each other, and the many differences between the models, including but not limited to the update intervals of their respective atmospheric forcing fields, cannot be easily reconciled.  Although ``apples-to-apples" comparisons are not feasible, we still believe that much can be learned from comparing their outputs to observations and to each other. 

\subsection{HYCOM simulation}
 
\begin{figure}
\noindent\includegraphics[width=0.9\textwidth]{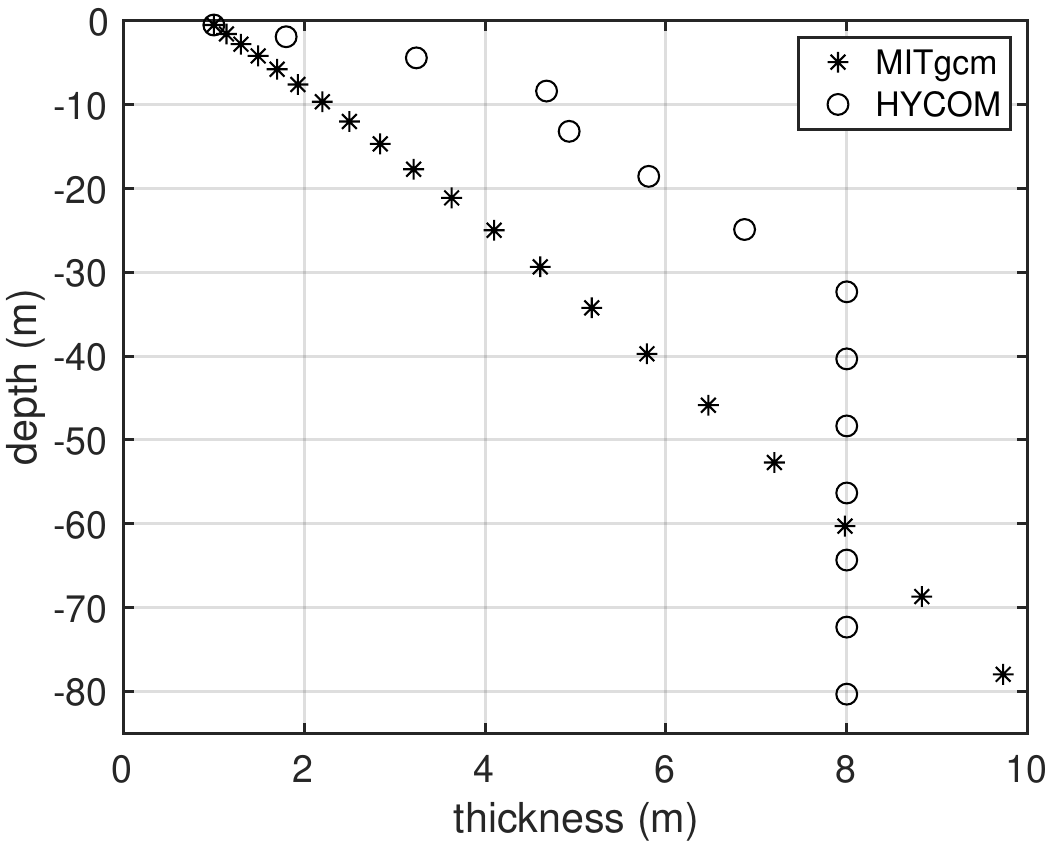}
\caption{Near-surface vertical discretization used in the HYCOM and MITgcm LLC4320 simulations.  The plot displays the grid cell thicknesses as a function of the midpoint depths of each grid cell.}
\label{fig:depths}
\end{figure}

The global HYCOM simulation employs nominal 1/25$^{\circ}$ horizontal grid spacing and 41 hybrid layers \cite{Bleck2002} covering the vertical direction.  HYCOM employs terrain-following coordinates in shallow waters, and isopycnal coordinates in the subsurface open-ocean.  In the near-surface open-ocean, the uppermost 14 layers are in z-mode, with seven z-levels, having spacing ranging from 1.00 to 6.87 m, in the uppermost 30 m (Figure \ref{fig:depths}).  The HYCOM ``0 m" (surface) results actually represent the mid-point of the uppermost layer.  The HYCOM ``15 m'' results represent interpolations to 15 m.  The interpolation largely reflects results from the HYCOM level at 13.185 m, which lies in between adjacent levels at 8.38 and 18.55 m.  The latter three depth values represent mid-depth points of the respective vertical layers in which they lie.  HYCOM employs the widely used K-Profile Parameterization (KPP) scheme \cite{LargeMcWilliamsDoney1994} for vertical mixing.  Atmospheric forcing fields from the U.S. Navy Global Environmental Model \cite<NAVGEM;>{Hogan2014} are applied every three hours and converted to surface fluxes using the bulk formulae of \citeA{Kara2000}.  HYCOM uses relative winds (10 m wind speed minus ocean surface current speed) to compute wind stress.  We use a 360-day record of hourly snapshots of surface and 15-meter horizontal velocity fields, starting on 1 January 2014, and produced with a 75 second baroclinic timestep.

HYCOM’s tidal forcing includes the two largest diurnal components (K$_{1}$ and O$_{1}$) and the three largest semidiurnal components (M$_{2}$, S$_{2}$, and N$_{2}$).  These five constituents capture most of the tidal variability in the oceans.  Collectively, they dissipate 3.437 TW \cite{EgbertRay2003}, while the largest eight tidal constituents (the five above, plus P$_{1}$, Q$_{1}$, and K$_{2}$) dissipate 3.508 TW, only 2$\%$ larger than the dissipation rate of the five constituents included in HYCOM.  The Self-attraction And Loading \cite<SAL;>{Hendershott1972,Ray1998} term is taken from the altimetry-constrained TPXO8 barotropic tide model \cite{Egbert1994,Egbert2002}.  The HYCOM parameterized topographic wave drag scheme, taken from \citeA{Jayne2001}, is tuned to minimize the M$_{2}$ surface elevation errors with respect to TPXO8.  The HYCOM simulation analyzed here employs a perturbation \cite{Ngodock2016}, generated with an Augmented State Ensemble Kalman Filter (ASEnKF), that reduces the area-weighted error of the surface tidal elevations, computed with respect to TPXO8 in waters deeper than 1000 m and latitudes equatorward of 66$^{\circ}$, to 2.6 cm.  Other than this indirect and offline use of data assimilation techniques, the HYCOM simulation is free-running (non-assimilative).

\subsection{MITgcm LLC4320 simulation}

LLC4320 is a global MITgcm free-running (non-assimilative) simulation with nominal 1/48$^{\circ}$ horizontal grid spacing and 90 vertical z-levels.  There are 13 z-levels in the uppermost 30 m, with thickness ranging from 1.0 to 4.6 m (Figure \ref{fig:depths}).  As with HYCOM, the MITgcm LLC4320 ``0 m" (surface) results represent the mid-point of the uppermost level.  The LLC4320 ``15-m'' velocities are taken from the 9th grid cell from the surface, which spans 13.26 to 16.1 m depth.  LLC4320 also employs the KPP vertical mixing scheme.  LLC4320 is forced by the full luni-solar astronomical tidal potential and by six-hourly atmospheric fields from the 0.14$^{\circ}$ European Centre for Medium-Range Weather Forecasting (ECMWF) operational model analysis, starting in 2011. The ECMWF atmospheric fields are converted to surface fluxes using the bulk formulae of \citeA{Large2004} and the dynamic/thermodynamic sea ice model of \citeA{Losch2010}.  The computation of air-sea momentum exchange is based on surface-relative wind, that is, ocean and sea-ice surface velocity are subtracted from 10-m atmospheric wind prior to computing surface stress.  The LLC4320 integration time step is 25 s.  We use a year-long record of near-surface horizontal velocity fields, saved as hourly snapshots, beginning on 12 November 2011.

We have found that the tidal forcing in the LLC4320 simulation has inadvertently been overestimated by a factor of 1.1121 while the SAL term has been omitted.  The effects of these tidal forcing errors on LLC4320 tides will be reported on in detail elsewhere.  In the results shown here, LLC4320 KE in semidiurnal and diurnal bands is not corrected for these errors.  We note again that the LLC4320 simulation does not employ a parameterized topographic wave drag.  

\subsection{Vertical momentum forcing and mixing near the surface}

Here we discuss the near-surface vertical momentum mixing parameterizations in the HYCOM and LLC4320 simulations.  If momentum is input at the surface of the ocean, differences between KE at 0 and 15 m will depend on how momentum is mixed downward.  In models the downward mixing of momentum is controlled by the vertical mixing scheme employed, in particular, the assumed vertical profile of vertical viscosity. Both the HYCOM and LLC4320 simulations employ the KPP scheme to set vertical viscosity but there are a few differences in the details of KPP implementation.  In particular, the LLC4320 simulation has finer vertical discretization near the surface (13 levels in the uppermost 30 m vs 7 for HYCOM, see Figure \ref{fig:depths}); the reference buoyancy and reference velocity used to compute KPP boundary layer depth, $B_r$ and $V_r$, respectively, in Equation 21 of \citeA{LargeMcWilliamsDoney1994} is computed in the surface 1-m-thick level for LLC4320 and in the top 10\% of the boundary layer depth in HYCOM; and the critical Richardson number $R_{ic}$ is 0.356 in LLC4320 while it is 0.25 in HYCOM.

Within the KPP scheme \cite{LargeMcWilliamsDoney1994}, the vertical viscosity is a cubic profile of depth, which yields increasing viscosity values from a finite surface value to a sub-surface maximum and then decreasing values below this maximum, down to a background viscosity at the base of the KPP boundary layer.  When such a cubic profile, or a linear profile, is applied to the wind-driven Ekman momentum equation for the surface boundary layer, this results in a frequency-dependent shear.  This shear is minimum at the inertial frequency and increases away from the inertial frequency \cite{elipot2006}.  This theoretical framework is useful for understanding how the locally-wind-driven component of oceanic currents \cite{LillyandElipot2021}, ranging from the inertial frequency to low-frequency Ekman motions, can be sheared in the upper 15 m of the ocean.  In addition, upper-ocean stratification modulates the ultimate penetration depth of wind momentum \cite{LargeCrawford1995,CrawfordLarge1996,Elipot2009,Dohan2011,LillyandElipot2021}.  

Some guidance regarding the sensitivity of the KPP scheme to the various empirical parameters that it contains is available in \citeA{VanRoekel2018} and \citeA{Li2019}.  A comparison of the performance of KPP relative to other widely used vertical mixing schemes in a HYCOM simulation can be found in \citeA{Pottapinjara2022}.  They showed that some schemes can be better than others at some locations and some seasons but that there is no universally “best” answer.  Quantifying the exact impact of the HYCOM and LLC4320 KPP implementation details would require computationally expensive sensitivity experiments, which are  beyond the scope of the present study.

\subsection{Ocean Surface Drifters}

In situ estimates of ocean near-surface velocities are obtained from the NOAA's GDP \cite{LumpkinOzgokmenCenturioni2017} which maintains an array of surface drifting buoys, currently tracked by the Global Positioning System and previously by the Argos system.  We use version 1.04c of the hourly high-frequency dataset \cite{Elipot2016} containing 17,324 individual surface drifter trajectories from October 1987 to June 2020, totalling $\sim$166M estimates of hourly positions and velocities. The spatial coverage of the drifter dataset is global, yet inhomogeneous with higher data density in convergence zones in the middle of ocean gyres, and sparse observations at the equator due to Ekman divergence, which tends to disperse drifters away \cite{Elipot2016}.  The usage of more than 30 years of drifter data, in contrast to the one year of output from each of the HYCOM and MITgcm LLC4320 simulations, is another source of difference that may affect the results presented here.  Until about 2006, the drifter dataset was rather sparse, such that only a few patches of ocean were covered.  Even in recent years, wide spatial gaps are prevalent (see, for instance, Figure 1c in \citeA{Yu2019} and note that the y-axis scale is logarithmic).  Therefore it is necessary to use all years of drifter observations in order to obtain global coverage that approximates what one obtains from just one year of output from a high-resolution model.  

Drifter data are intrinsically Lagrangian, in contrast to the Eulerian model outputs to which they are compared. Nevertheless, segments of velocity time series from drifters can be used to estimate spectra and kinetic energy locally, yet keeping in mind that Lagrangian sampling leads to spectral smearing as drifters convolve spatial and temporal oceanic variability \cite{Yu2019,ZaronElipot2021}.  For example, Lagrangian spectra have lower and wider tidal peaks, which do not stand above the background as much as peaks in Eulerian spectra do.  In addition, tidal lines in both Lagrangian and Eulerian spectra widen due to interaction with currents and eddies, which renders the tides less coherent or stationary \cite<e.g.,>{RayZaron2011,Shriver2014,Zaron2014,Ponte2015,Kerry2016,Buijsman2017,Savage2017a,Zaron2017,Nelson2019}.

We utilize both drogued and undrogued drifter data, in waters deeper than 500 m.  Because the spatial and temporal distributions of the hourly drifter dataset are inhomogeneous (see for example Figure 1 of \citeA{Yu2019} and Figure 1 of \citeA{elipot2020measuring}), we find it necessary to consider the entire dataset to properly define a global mean velocity field upon which local variance estimates depend. Examination of the drifter data geographical density per calendar month (not shown) do not indicate seasonal bias. Drogued drifter displacements, which comprise 48\% of the trajectories in the hourly dataset version 1.04c, are expected to be representative of ocean velocity at 15-m depth with an estimated erroneous slip of the water past the drifter of 0.7 cm~s$^{-1}$ downwind per 10 m~s$^{-1}$ wind speed \cite{niiler1995wind}.  Undrogued drifters, which comprise 52\% of the dataset, are expected to represent ocean velocities at the surface (0 m), but with a slip of an order of magnitude larger compared to drogued drifters [8.6 cm~s$^{-1}$ per 10 m~s$^{-1}$ wind speed, \cite{LumpkinPazos2007Measuring}].  As such, undrogued drifter observations likely exhibit larger downwind velocity errors but these are yet to be comprehensively distinguished from real oceanic processes. For example, locally wind-driven velocities at the surface are more energetic than at 15-m depth because of vertical shear at a broad range of frequencies through Ekman dynamics \cite{Elipot2009,LillyandElipot2021}, or through surface gravity wave processes and their associated Stokes drift \cite{Polton2005}. Yet, as will be seen throughout this paper, undrogued drifters qualitatively capture the same KE features as drogued drifters.  A correction for the wind slip or an adequate assessment of its magnitude would need to be informed by an unknown dependency on frequency and to take into account the entire frequency spectrum of the observable wind forcing. In addition, typical estimation errors for the hourly drifter velocity estimates \cite{Elipot2016} are between 2 and 5 cm~s$^{-1}$ \cite<see Figure S2 of>{Yu2019} with unknown frequency distribution. As such, a comprehensive assessment of the velocity errors from drogued and undrogued drifters, and how these errors affect signal-to-noise ratios and kinetic energy estimates in various frequency bands, is beyond the scope of this study.  

\subsection{Analysis methods}

Following \citeA{Yu2019}, velocity variance is estimated and interpreted as KE; no factor of 1/2 is included in the KE calculations. In order to study the distribution of KE as a function of time scale, we rely on estimating frequency rotary spectra \cite{Gonella1972,Mooers1973} of velocity time series. Such spectra allow us to decompose velocity variance as a function of frequency, and to separate clockwise versus counterclockwise variances in order to distinguish anticyclonic energy from cyclonic energy.

To generally estimate velocity rotary spectra from models and drifters, complex velocity time series $u+iv$, where $u$ and $v$ respectively denote the zonal and meridional velocity components, are split into 60-day segments overlapping by 50\%, detrended, and individually multiplied by a normalized Hann window to reduce spectral leakage. The discrete Fourier transform is then computed for all windowed segments and multiplied by their complex conjugates in order to obtain initial spectral estimates. To reduce computational time, model outputs are first subsampled on a 1/4$^{\circ}$ grid before estimating spectra. Since we are using one-year model integrations, we can split each model velocity time series at each selected Eulerian grid point, and at each model depth (0 and 15 m), into 11 individual overlapping 60-day segments, discarding only a few days of data.  Each segment is used to compute an initial spectral estimate, and the 11 resulting estimates are averaged to obtain a spectrum representative of the velocity variance for time scales of 60 days and shorter, at each individual location. This method should average the seasonality across the one-year model integrations, if present. These time-averaged spectra are further averaged in $1^\circ \times 1^\circ$ spatial bins before conducting subsequent analyses. Spatial averages are only computed when more than 50$\%$ of the points in a 1$^{\circ}$ by 1$^{\circ}$ bin are deeper than 500 m, and shallower gridpoints are discarded in the computation. For drifters, we divide up individual trajectories in as many 60-day overlapping segments as possible, discarding trailing data. For the dataset version 1.04c, we can obtain 174,803 60-day segments, 52,710 of which are from drifters with drogue on (drogued) and 108,861 are from drifters with drogue off (undrogued). This leaves 13,223 segments which are partly drogued and undrogued and are not considered to compute spectra. 

In the next Results section, we consider zonally-averaged spectra (Figure~\ref{fig:frequency_latitude_spectra}). For the models, we simply calculate the average of all spectra within a given 1$^\circ$ latitude zonal bin. For the drifters, we average together the individual spectra for which the average latitudes of the corresponding segments are located within a given 1$^\circ$ latitude zonal bin. We also consider globally averaged spectra (Figure~\ref{fig:globalavg_spectra}). For the models, we calculate a weighted average of all the $1^\circ \times 1^\circ$ spectra, where the weights are proportional to the surface area of the bins. For the drifters, we calculate weighted averages of individual initial spectra, separately for the drogued and undrogued drifters, where the weights are proportional to the cosine of the mean latitude of each corresponding segment. For both model and drifter globally-averaged spectra, we make sure to average spectra from anticyclonic frequencies together, and spectra from cyclonic frequencies together, the sign of which depends on the hemisphere of the data. For the drifters, we only consider trajectory segments for which the median of depths traversed along a segment is deeper than 500 m. Depths along drifter trajectories are obtained by interpolating the ETOPO1 global relief model \cite{amante2009etopo1}.

In order to produce maps of KE in specific frequency bands, we proceed as in \citeA{Yu2019} for the models: the $1^\circ \times 1^\circ$ gridded rotary spectra are integrated over various frequency bands: semidiurnal ($\pm [1.9,2.1]$ cpd), diurnal ($\pm [0.9,1.1]$ cpd), high frequency ($> 0.5$ cpd and $<-0.5$ cpd), and near-inertial ($\pm [0.9,1.1] f$, where $f$ is the Coriolis frequency).  Low-frequency KE is taken as total KE, computed as the time-mean of the squares of the zonal velocity time series plus the time-mean of the squares of the meridional velocity time series, minus high-frequency KE, the latter computed from the spectra. 

To produce maps of KE from Lagrangian drifters in a specific frequency band, we do not integrate the drifter spectra as for the Eulerian models. Instead, for each band, we first apply a filter (bandpass or lowpass as needed) to the drifter velocity time series.  In the near-inertial band, we apply a bandpass filter akin to the method of complex demodulation. The details of the bandpass filtering of the drifter data are given in the Appendix.  Second, following bandpass filtering, we compute the variance of all filtered drifter velocities in $1^{\circ} \times 1^{\circ}$ spatial bins, independently of their drifter of origin. This new method based upon filtering differs from the study of \citeA{Yu2019} in which band-specific variance estimates were calculated by firstly integrating the spectra of 60-day segment time series, secondly assigning the mean geographical positions of the segments to these estimates, and thirdly averaging them in 1$^{\circ}$ spatial bins to produce maps.  The method employed in \citeA{Yu2019} yields results similar to the ones presented here. However, our new method yields better spatial coverage in the drifter maps, especially near the equator, as well as slightly higher correlations between the drifter and model maps.  
Caution is warranted when interpreting the results for the near-inertial band.  First, the near-inertial band as defined above covers only the ``local'' near-inertial KE.  Near-inertial motions, such as low-mode internal tides, can propagate over long distances \cite{Alford2003a,Simmons2012}, and in such cases their frequency is no longer equal to the local value of $f$.  Second, our analysis, like that of \citeA{Yu2019}, does not distinguish between near-inertial and diurnal motions where the definitions of these bands overlap, namely, within 24.1-37.5$^{\circ}$ of latitude. 

We compare KE maps between depth levels by considering a proxy of vertical structure, namely the ratio of undrogued KE divided by the sum of undrogued and drogued KE for the drifters, and the ratio of 0 m KE divided by the sum of 0 m and 15 m KE for the models. This ratio statistic is preferable to a straightforward ratio of 0m to 15 m because it has an upper bound value of one when the KE at 15 m depth is zero. The ratio statistic is equal to 0.5 when the KE at 15 m is equal to the KE at the surface, greater than 0.5 when the KE at 15 m is less than at the surface, and less than 0.5 when the 15 m KE is larger.  

Our comparisons of KE and ratio maps originating from drifters and models include Pearson correlation coefficients computed between maps. The statistical significance of each correlation estimate is assessed by drawing 1000 bootstrap sample pairs from which the standard error of the correlation is estimated. Using this method all non-zero correlations are deemed significant. 

Finally, zonal averages of drifter and model KE maps and their ratio statistics are computed.  Our plots of zonal averages display mean values and the standard error of the mean values computed over latitude bands.

\section{Results}

\begin{figure}
\noindent\includegraphics[width=\textwidth]{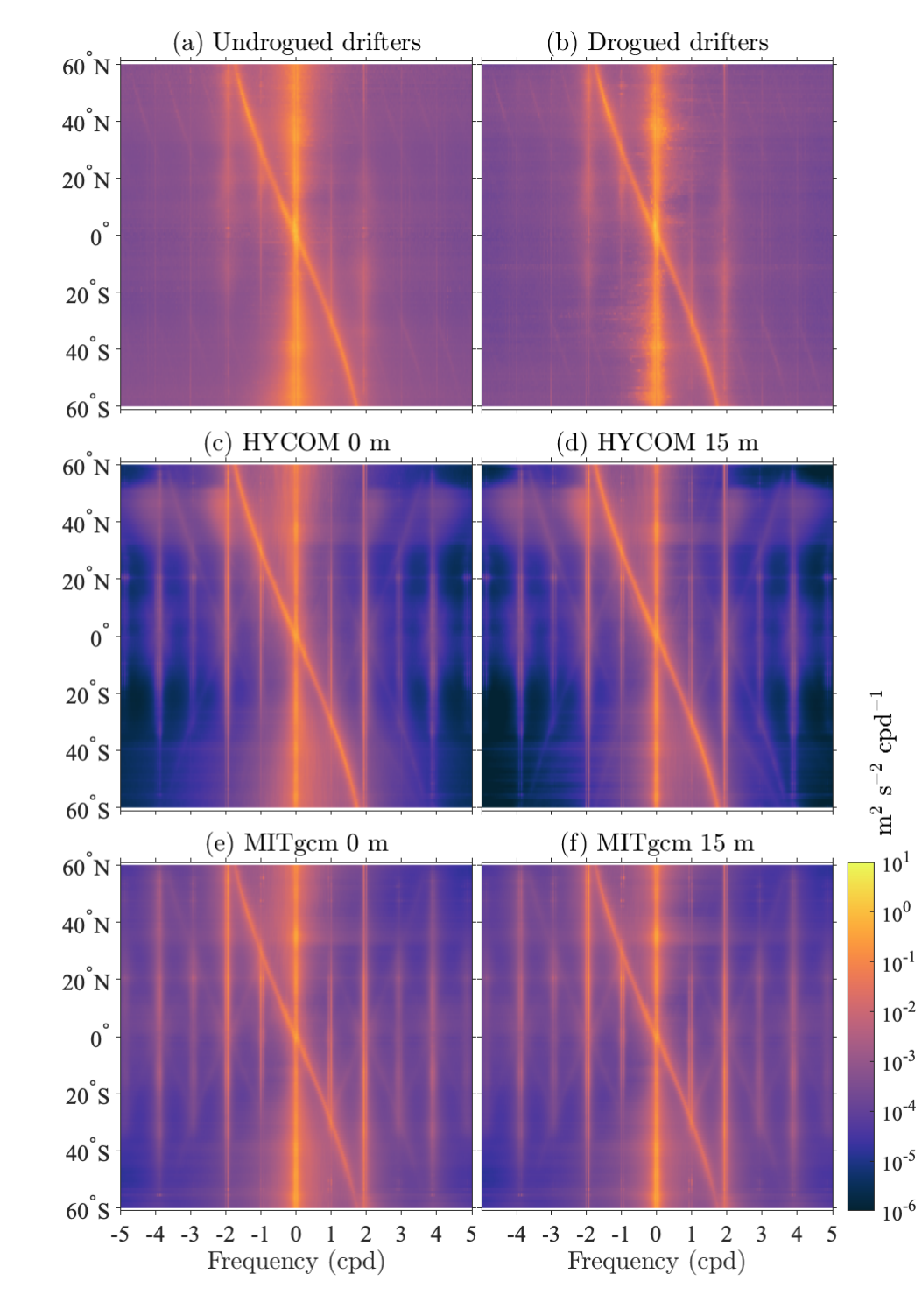}
\caption{Zonally averaged rotary spectra of KE in 1$^{\circ}$ latitude bins between 60$^{\circ}$S and 60$^{\circ}$N for undrogued and drogued drifters (panels a and b), for HYCOM at 0 m and 15 m (panels c and d), and for MITgcm LLC4320 at 0 m and 15 m (panels e and f). Only frequencies between -5 and 5 cycle per day (cpd) are displayed. The frequency resolution is 1/60 cpd. The common decimal logarithmic color scale is displayed at the bottom right.}
\label{fig:frequency_latitude_spectra}
\end{figure}

Zonally-averaged rotary spectra for undrogued and drogued drifters, and at 0 m and 15 m in the models, visualized as a function of frequency and latitude, display relatively similar KE levels and structures (Figure \ref{fig:frequency_latitude_spectra}).  Peaks corresponding to diurnal frequencies (near $\pm$ 1 cpd) and semidiurnal frequencies (near $\pm$ 2 cpd), and a large low-frequency continuum (near 0 cpd) are evident in all six subplots.  Ridges of near-inertial energy, which follow the negative of the Coriolis frequency, from ~$\sim$ -1.7 cpd at 60$^{\circ}$N to $\sim$ 1.7 cpd at 60$^{\circ}$S, are also evident in all six subplots of Figure \ref{fig:frequency_latitude_spectra}. Near 30$^{\circ}$ latitude, the inertial ridge overlaps with the diurnal peaks in the anticyclonic domain (at negative frequencies in the northern hemisphere, and positive frequencies in the southern hemisphere). The low-frequency peak, around zero cycles per day, is broader in frequency in the drifters than in the models. The models exhibit semidiurnal tidal peaks that rise above the background more dramatically than the peaks in the drifter spectra.  The models also display a ``peaky'' or ``picket fence'' distribution of energy at high-frequency tidal harmonics, with little energy in between the harmonics, as noted in previous studies \cite<e.g.,>{Muller2015,Savage2017b}.  In contrast, the drifter tidal peaks do not stand out above the background as strongly as in the models, as discussed earlier and in \citeA{ZaronElipot2021}. 

Another faint but striking feature of these latitude-frequency rotary spectra is the presence of translated images of the near-inertial ridge at $-f \pm 1,2,3$ cpd in both drifter and model spectra. The model spectra additionally exhibit mirror images of the inertial ridge at $f \pm 0,1,2,3$ cpd. These features were previously seen in a similar figure in \citeA{Yu2019} for drifters and MITgcm LLC4320, and for drifter data only in \citeA{Elipot2016} up to 12 cpd. \citeA{Elipot2016} speculated that these features may be representative of small departures from circular geometry for inertial oscillations, or of triad interactions between near-inertial waves and internal waves at tidal frequencies, but also noted that their magnitudes depended to some extent upon the processing applied to the drifter data to obtain hourly velocity estimates. The fact that these features are also observed in the models suggest that they might be  representative of real, as yet unexplained, oceanic processes.   

\begin{figure}
\noindent\includegraphics[width=\textwidth]{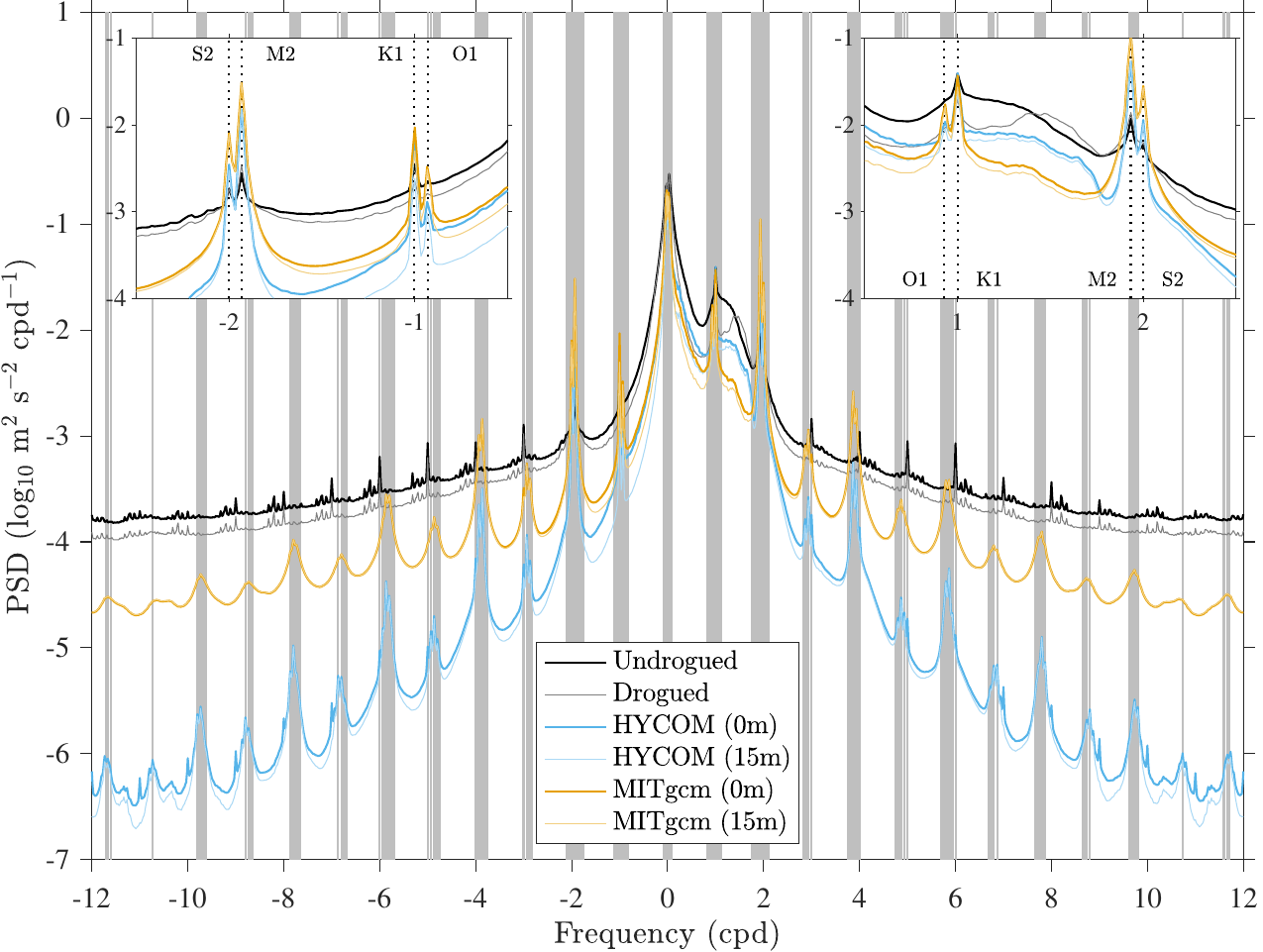}
\caption{Globally averaged rotary spectra of KE from undrogued and drogued drifters (black and gray curves, respectively), from 0 m and 15 m HYCOM (thick and thin cyan curves, respectively), and from 0 m and 15 m MITgcm LLC4320 (thick and thin gold curves, respectively). Anticyclonic and cyclonic frequencies are assigned to positive and negative frequencies, respectively. The vertical gray lines indicate 145 tidal frequencies in both positive and negative frequency domains. The two insets show the same spectra but with a focus on diurnal and semidiurnal anticyclonic frequencies (right) and cyclonic frequencies (left). The vertical dotted lines in the insets show the O$_{1}$, K$_{1}$, M$_{2}$, and S$_{2}$ frequencies.}
\label{fig:globalavg_spectra}
\end{figure}

The relative energy levels in the different bands are more easily compared in globally averaged rotary spectra, for which the anticyclonic domain is assigned here to positive frequencies and the cyclonic domain to negative frequencies, using a  southern hemisphere convention (Figure \ref{fig:globalavg_spectra}).  The highest energy levels, associated with large-scale currents, mesoscale eddies, and Ekman flows, are seen in high and wide peaks at low frequencies (around zero).  Whereas the near-inertial anticyclonic ridge is clearly visible in frequency-latitude spectra (Figure \ref{fig:frequency_latitude_spectra}), in globally averaged spectra, that ridge is instead spread unevenly between 0 and ~1.7 cpd in the anticyclonic frequency domain, weighted by the latitudinal distribution of the spectral estimates. As a result, energy levels are generally higher below the semidiurnal frequency band in the anticyclonic domain than in the cyclonic domain.  Semidiurnal and diurnal peaks are clearly visible in both the model and drifter spectra but rise above the background much less in the drifter spectra than in the model spectra, as is made clear in the insets of Figure \ref{fig:globalavg_spectra}.  The insets illustrate the wider and lower semidiurnal tidal peaks in Lagrangian spectra in comparison to Eulerian spectra.  As noted earlier, in the higher frequency part (e.g., $> 2$ cpd) of the IGW continuum, spectra are much peakier in the models, especially HYCOM.  The IGW continuum falls off more steeply in HYCOM than in MITgcm LLC4320, because of the lower resolution (coarser grid spacings) in HYCOM.  The continuum is more elevated and smoother in the drifter spectra, and displays a noise floor at 12 cpd which is about one decade higher than MITgcm LLC4320, and more than two decades higher than HYCOM.  This noise floor depends on the tracking-system for drifters.  As shown by \citeA{Yu2019}, when only GPS-tracked drifters are considered instead of Argos-tracked drifters \cite{Elipot2016}, the spectral level and spectral slope at the highest frequencies for drifter spectra is in approximate agreement with MITgcm LLC4320 (see Figure 2 of \citeA{Yu2019}). Many small but distinct spectral peaks that do not necessarily correspond to tidal constituents are seen in the drifter spectra for frequencies higher than 4 cpd. The amplitude and frequency of these peaks depend also on the drifter tracking system (see Figure 2 of \citeA{Yu2019}) and may be artifacts of the estimation methods for drifter position and velocities \cite{Elipot2016}.

\subsection{Low-frequency KE}

\begin{figure}
\centering
\noindent\includegraphics[width=\textwidth]{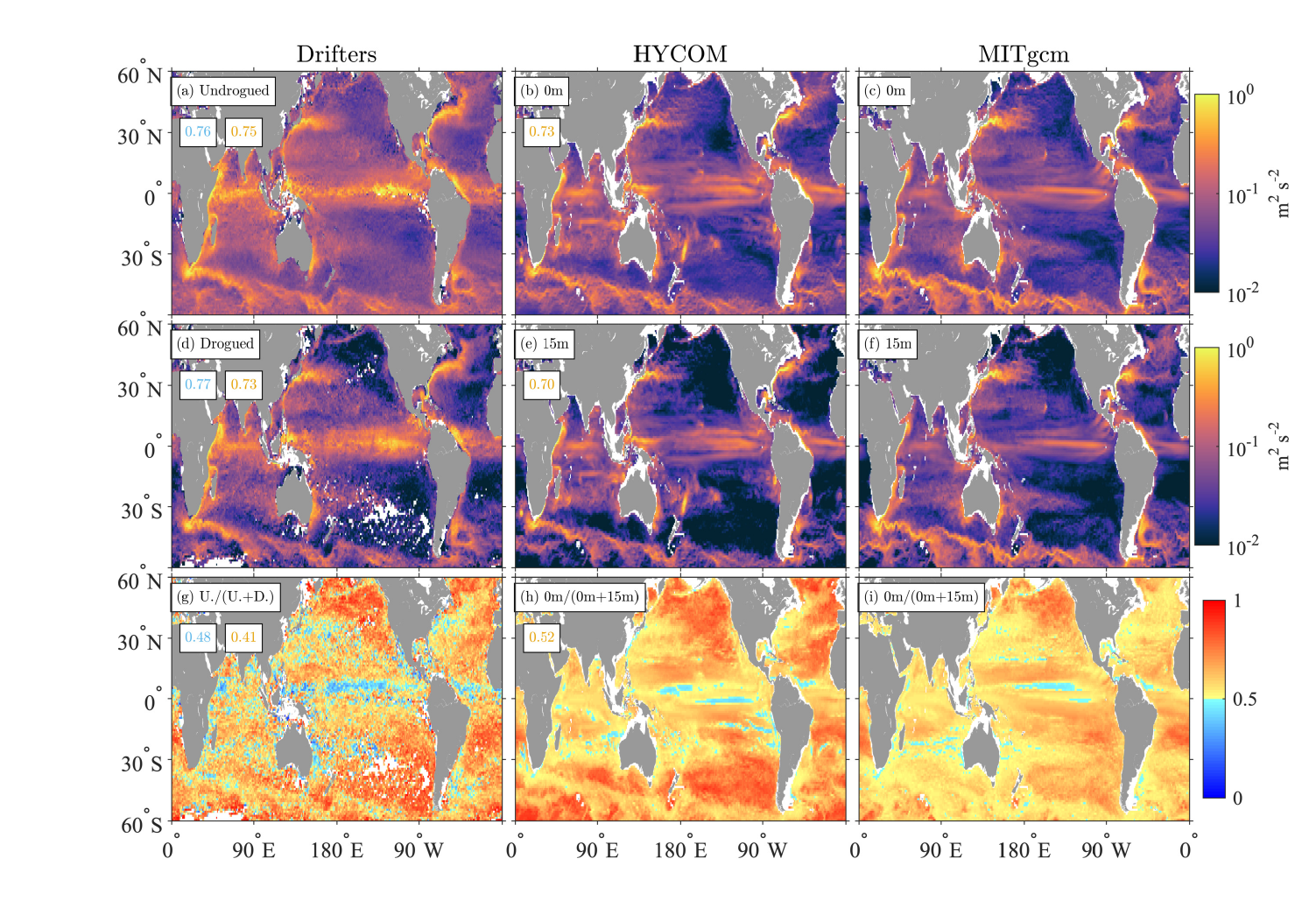}
\caption{Global maps of low-frequency ($>$ -0.5 cpd and $<$ 0.5 cpd) kinetic energy (KE) from undrogued and drogued drifters (panels a and d), from HYCOM at 0 m and 15 m (panels b and e), and from MITgcm LLC4320 at 0 m and 15 m (panels c and f). The ratios of undrogued KE/(undrogued KE + drogued KE) for the drifters, and 0 m KE/(0 m KE + 15 m KE) for the HYCOM and MITgcm LLC4320 simulations, are shown in panels g, h, and i, respectively. The spatial correlations between the drifter maps and HYCOM and MITgcm LLC4320 maps are indicated in the upper left corners of panels a, d, and g, with HYCOM values in cyan to the left of MITgcm LLC4320 values in gold. The spatial correlations between HYCOM and MITgcm LLC4320 maps are indicated in the upper left corners of panels b, e, and h.}
\label{fig:lowf_maps}
\end{figure}

Global maps of low-frequency KE highlight well-known large-scale currents and the mesoscale eddies they spawn, both of which are dominant in equatorial regions, western boundary current regions such as the Gulf Stream and Kuroshio, and the Antarctic Circumpolar Current (Figure~\ref{fig:lowf_maps}, panels a-f). Ekman flows also contribute to these low-frequency KE patterns \cite{lumpkin2013global}.  At both depth levels, HYCOM overestimates the drifter observations in the near-equatorial southern Indian Ocean and in the Pacific off the Western coast of South America, while MITgcm LLC4320 overestimates the drifters in the eastern North Atlantic Ocean and parts of the Southern Ocean.  MITgcm LLC4320 also features an incorrect positioning of the Gulf Stream, which does not veer to the northeast as it does in HYCOM and the drifters.  The same Gulf Stream patterns were noted in surface velocities computed from satellite altimetry in \citeA{Luecke2020}. Despite these local differences, the spatial correlations between HYCOM and the drifters (0.76 and 0.77, for undrogued and drogued drifters, respectively) are comparable to, though slightly higher than, the correlations between MITgcm LLC4320 and the drifters (0.75 and 0.73, for undrogued and drogued drifters). 

Maps of the low-frequency band vertical structure proxy ratio (Figure~\ref{fig:lowf_maps}, panels g-i) display some clear similarities, despite the noisiness inherent in the drifter map.  The mid- to high-latitude northeast and southeast Pacific Ocean, for instance, display similarly large values of the ratio (exceeding 0.5) in all three maps. In some specific locations such as near the equator in the Pacific, values of the ratio statistic slightly less than 0.5 are seen in both the model ratio maps and the drifter map. This suggests that some robust features of the oceanic circulation, both observed and modeled, are such that KE increases from the surface to 15 m depth.  The spatial correlation between the model ratio maps (0.52) is higher than the spatial correlation between either model and the drifters.  The HYCOM correlation with the drifter ratio map (0.48) is higher than the MITgcm LLC4320 correlation with the drifters (0.41).  

\begin{figure}
\centering
\noindent\includegraphics[width=\textwidth]{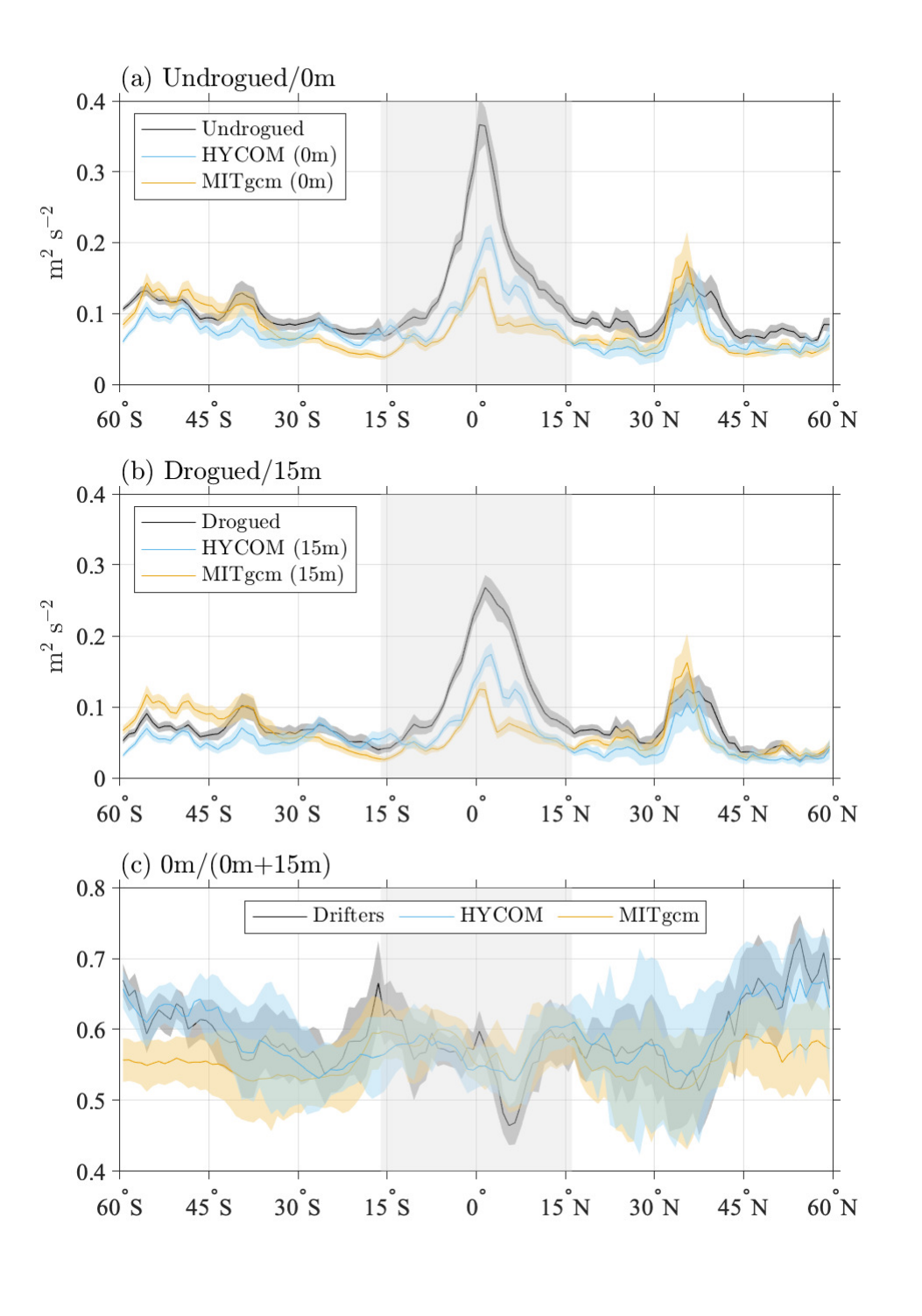}
\caption{Zonally averaged low-frequency ($>$ -0.5 cpd and $<$ 0.5 cpd) kinetic energy (KE) from (a) undrogued drifters and 0~m model levels, and (b) drogued drifters and 15~m model levels.  Zonal averages of undrogued KE/(undrogued KE + drogued KE) for the drifters, and 0 m KE/(0 m KE + 15 m KE) for HYCOM and MITgcm LLC4320, are shown in (c). The gray shaded region in each panel indicates the latitudes where the near-inertial and low-frequency bands exhibit some overlap (see Fig.~\ref{fig:frequency_latitude_spectra}). The shading around each curve corresponds to two standard errors of the calculated 1$^\circ$ zonal averages from the geographical maps.}
\label{fig:lowf_zavg}
\end{figure}

Zonally averaged low-frequency KE in both models is generally comparable to, but lower than, drifter KE (Figure~\ref{fig:lowf_zavg}, panels a and b).  The models and drifters all exhibit a peak of energy at the equator, but the peak values in HYCOM and especially MITgcm LLC4320 are too low, by a factor of about two.  This disagreement between models and drifters near the equator should be interpreted with some caution as the sampling density from drifters near the equator is relatively low \cite<see Figure 1(a) of>{Yu2019}. HYCOM KE is too low across the mid- to high-latitudes of the southern hemisphere.  With the exception of the energetic peak near 33$^{\circ}$N, over most mid- to high-latitudes in the northern hemisphere, both models are too low.

The zonally averaged low-frequency band vertical structure proxy ratio in the models tracks the zonally averaged ratio in the drifters reasonably well (to within error bars) over most latitudes (Figure~\ref{fig:lowf_zavg}c).  Maximum values of the ratio are seen at high latitudes in both hemispheres in the drifters and in HYCOM, whereas MITgcm LLC4320 does not track this high-latitude behavior.  The greater energy in the undrogued drifters over most latitudes (represented by values of the ratio greater than 0.5) may be due in part to their wind bias as noted earlier.  However, the relative closeness of this KE ratio in results from models, which do not suffer from a wind bias, to the drifter results suggests that wind bias alone is unlikely to account for all of the vertical structure seen in Figure~\ref{fig:lowf_zavg}c.  

\subsection{Near-inertial KE}

\begin{figure}
\centering
\noindent\includegraphics[width=\textwidth]{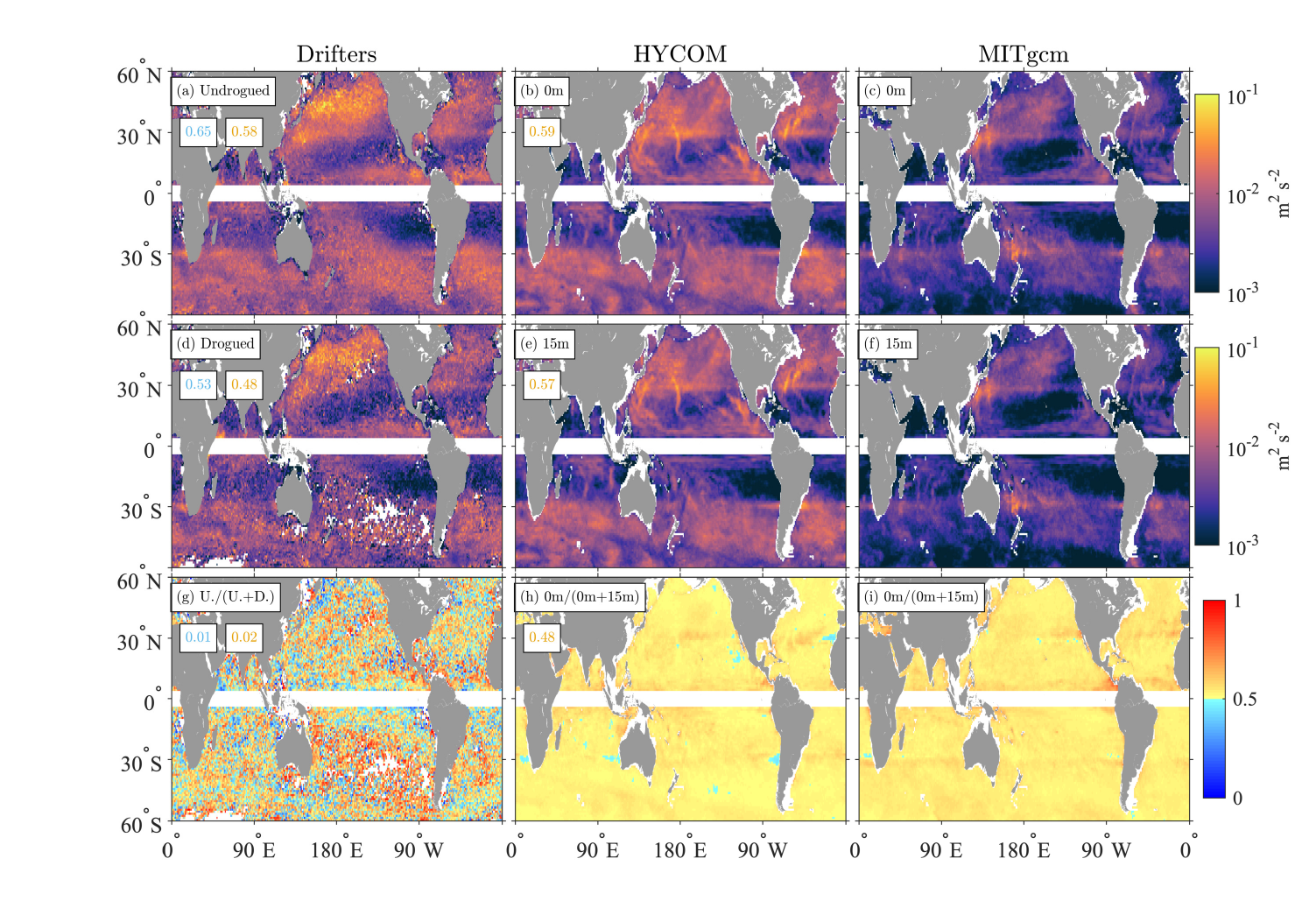}
\caption{Global maps of near-inertial ($\pm [0.9,1.1] f$) KE from undrogued and drogued drifters (panels a and d), from HYCOM at 0 m and 15 m (panels b and e), and from MITgcm LLC4320 at 0 m and 15 m (panels c and f). The ratios of undrogued KE/(undrogued KE + drogued KE) for the drifters, and 0 m KE/(0 m KE + 15 m KE) for the HYCOM and MITgcm LLC4320 simulations, are shown in panels g, h, and i, respectively.  The spatial correlations between the drifter maps and HYCOM and MITgcm LLC4320 maps are indicated in the upper left corners of panels a, d, and g, with HYCOM values in cyan to the left of MITgcm LLC4320 values in gold. The spatial correlations between HYCOM and MITgcm LLC4320 maps are indicated in the upper left corners of panels b, e, and h.}
\label{fig:near_inertial_maps}
\end{figure}

Near-inertial KE is generally larger in mid-to-high latitudes than it is near the equator for both observational and model results, with the largest values found in the North Pacific (Figure~\ref{fig:near_inertial_maps}, panels a-f), as shown in previous studies \cite<e.g.,>{Alford2003b,chaigneau2008global,elipot2010modification}.  Faint but distinct relative maxima of KE are evident in the maps for the models around 30$^\circ$ where the diurnal and near-inertial bands overlap.  MITgcm LLC4320 systematically underestimates the drifter results over most of the ocean, while HYCOM is in generally better agreement though it still underestimates the drifter results in the northern Pacific.  Near-inertial KE is conspicuously lacking in MITgcm LLC4320 over the Antarctic Circumpolar Current south of 45$^\circ$S.  In both models, near meridional streaks of near-inertial KE stand out, probably related to individual tropical cyclones and storms present in the model forcing fields within their respective integration years.  The drifter maps do not show such features which should be averaged out by the many years of drifter data used for this analysis.  The different forcing years of the models and the multi-year nature of the drifter dataset may explain why the spatial correlations between these maps at both levels are lower than seen in, for instance, the low-frequency band; the two model maps correlate at 0.59 at the surface and at 0.57 at 15 m depth. The drifter maps correlate with the model maps at approximately equivalent levels, albeit slightly higher in HYCOM compared to MITgcm LLC4320 (0.65 versus 0.58 at the surface, and 0.53 versus 0.48 at 15 m).

The maps of near-inertial band vertical structure proxy ratio for the two models (Figure~\ref{fig:near_inertial_maps}, panels h-i) appear similar but exhibit relatively low correlation between them (0.48). They both indicate that near-inertial KE is slightly larger at the surface compared to 15 m. Some unexplained discontinuities in these maps are noticeable near $30^\circ$ latitude.  The discontinuities may be due to overlapping dynamical processes of a different nature taking place there (tidal vs. wind-driven). The ratio map for the drifters (Figure~\ref{fig:near_inertial_maps}, panel g) is noisy, and exhibits near-zero correlations with the model ratio maps.

\begin{figure}
\centering
\noindent\includegraphics[width=\textwidth]{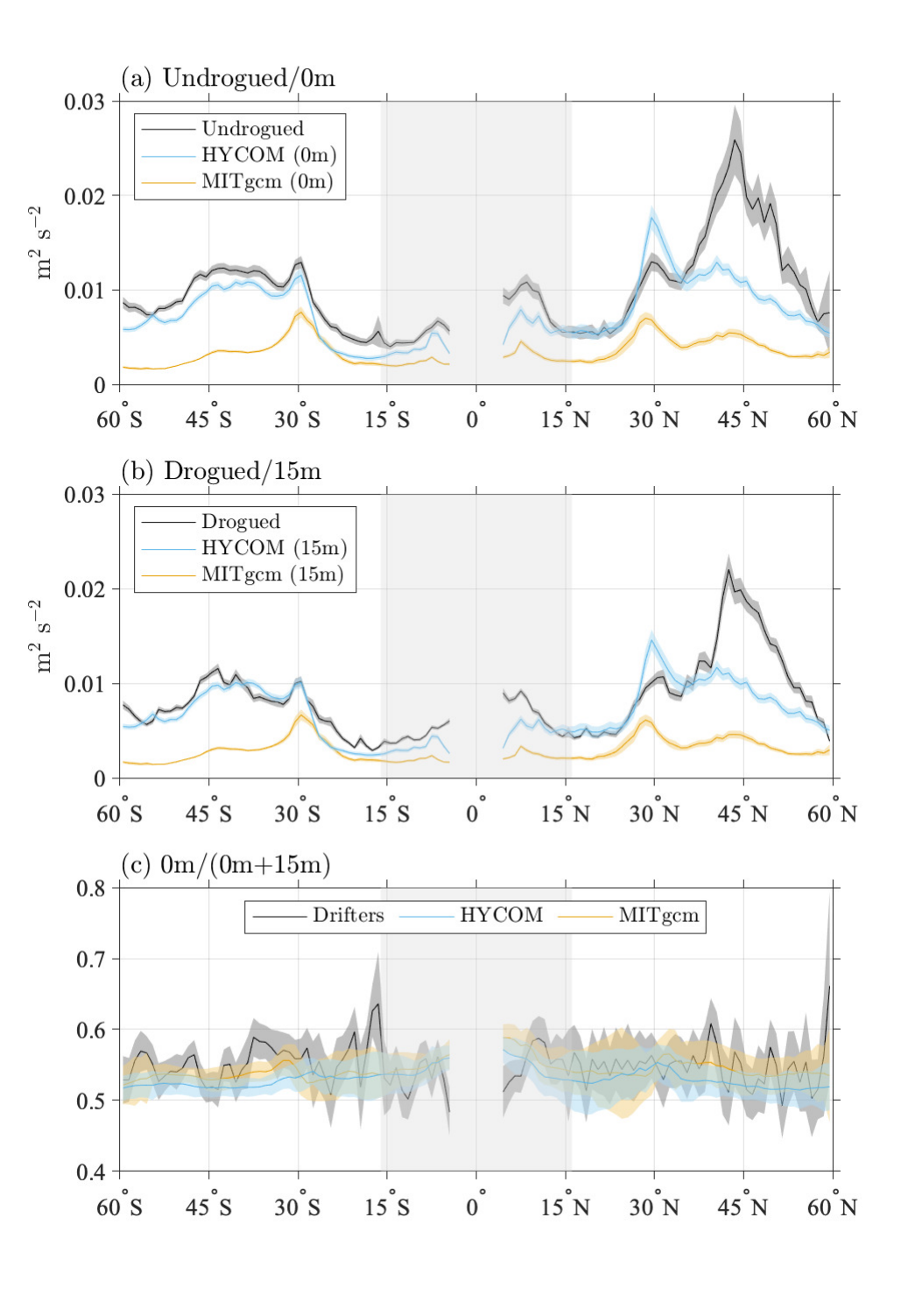}
\caption{Zonally averaged near-inertial ($\pm [0.9,1.1] f$) KE from (a) undrogued drifters and 0~m model levels, and (b) drogued drifters and 15~m model levels.  Zonal averages of undrogued KE/(undrogued KE + drogued KE) for the drifters, and 0 m KE/(0 m KE + 15m KE) for HYCOM and MITgcm LLC4320, are shown in (c).  The gray shaded region in each panel indicates the latitudes where the near-inertial and low-frequency bands exhibit some overlap (see Fig.~\ref{fig:frequency_latitude_spectra}). The shading around each curve corresponds to two standard errors of the calculated 1$^\circ$ zonal averages from the geographical maps.}
\label{fig:near_inertial_zavg}
\end{figure}

\begin{figure}
\centering
\noindent\includegraphics[width=\textwidth]{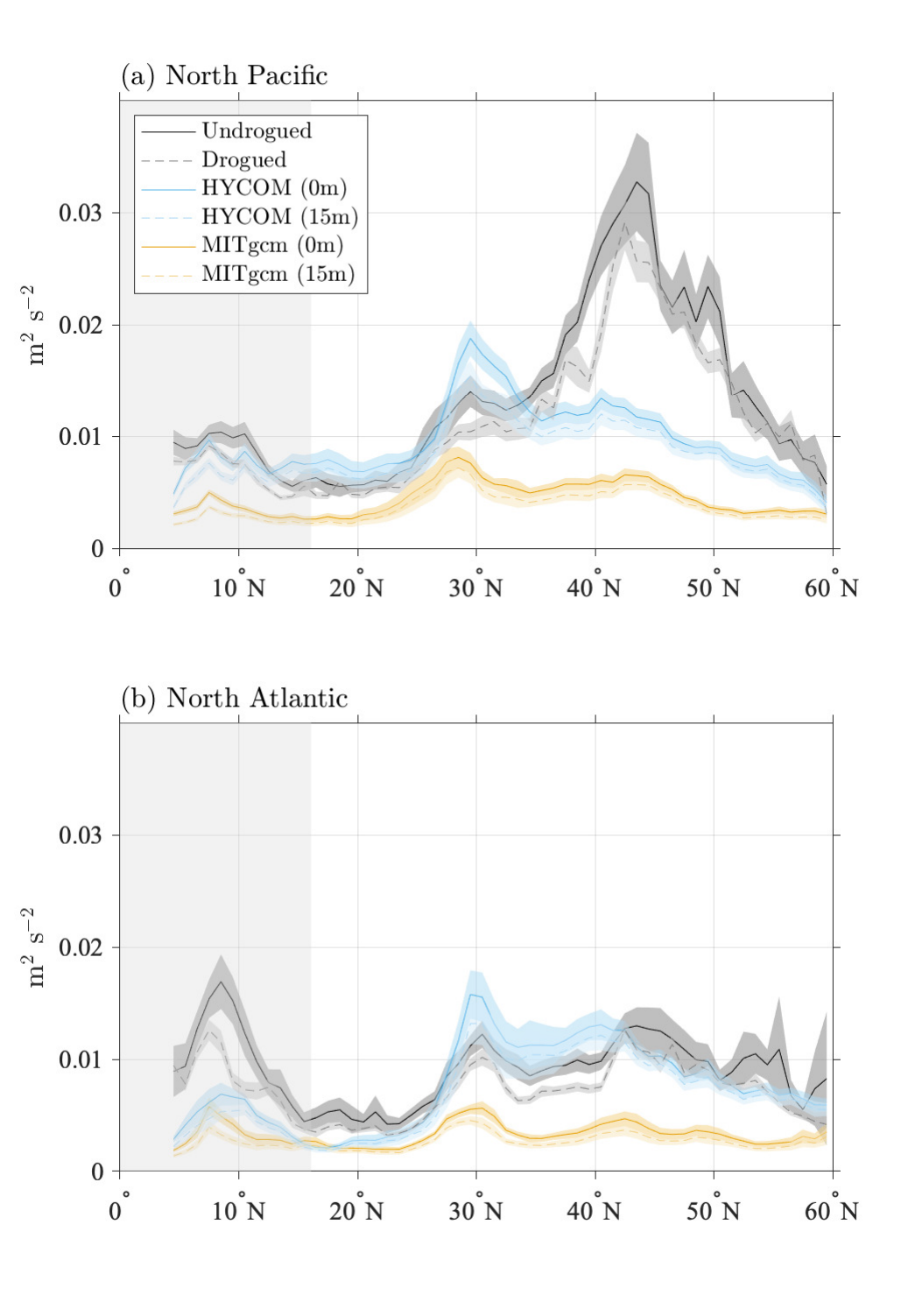}
\caption{Zonally averaged near-inertial KE, as in Fig.~\ref{fig:near_inertial_zavg}a and b, but with North Atlantic and North Pacific Ocean basins examined separately.  Note that in this figure, both undrogued/0 m and drogued/15 m results are displayed on both subplots.}
\label{fig:near_inertial_ocean_basin}
\end{figure}

In the zonal averages (Figure~\ref{fig:near_inertial_zavg}), as anticipated from the global maps, near-inertial KE is significantly higher in HYCOM than in MITgcm LLC4320.  Near-inertial KE in HYCOM follows the drifters relatively well in the northern hemisphere between about 10-35$^{\circ}$N and in the southern hemisphere.  In the southern hemisphere, at the surface, the undrogued drifter near-inertial KE is however systematically slightly higher than HYCOM, perhaps because of underestimated windage of the undrogued drifters within strong wind environments.  In contrast, at 15-m depth, HYCOM is in closer agreement with the drogued drifters between about 30-60$^{\circ}$S.  At latitudes between about 35-60$^{\circ}$N, HYCOM KE values are closer to the drifters than MITgcm LLC4320 values are.  However, over these latitudes HYCOM near-inertial KE is still substantially lower than drifter near-inertial KE. Separation of the zonal averages into basins demonstrates that the main cause of the discrepancy is located in the North Pacific Ocean (Figure~\ref{fig:near_inertial_ocean_basin}).  

The zonally averaged vertical structure proxy ratio for the near-inertial band is still noisy in the drifters,  but is higher than 0.5 over most latitudes (Figure~\ref{fig:near_inertial_zavg}c). Over most latitudes, the model zonally averaged vertical structure proxy ratios follow the drifter ratios relatively well, within error bars. The model zonally averaged ratios are slightly larger than 0.5.  Values between about 0.50-0.55, for instance, are typical.  

\subsection{Diurnal KE}

\begin{figure}
\centering
\noindent\includegraphics[width=\textwidth]{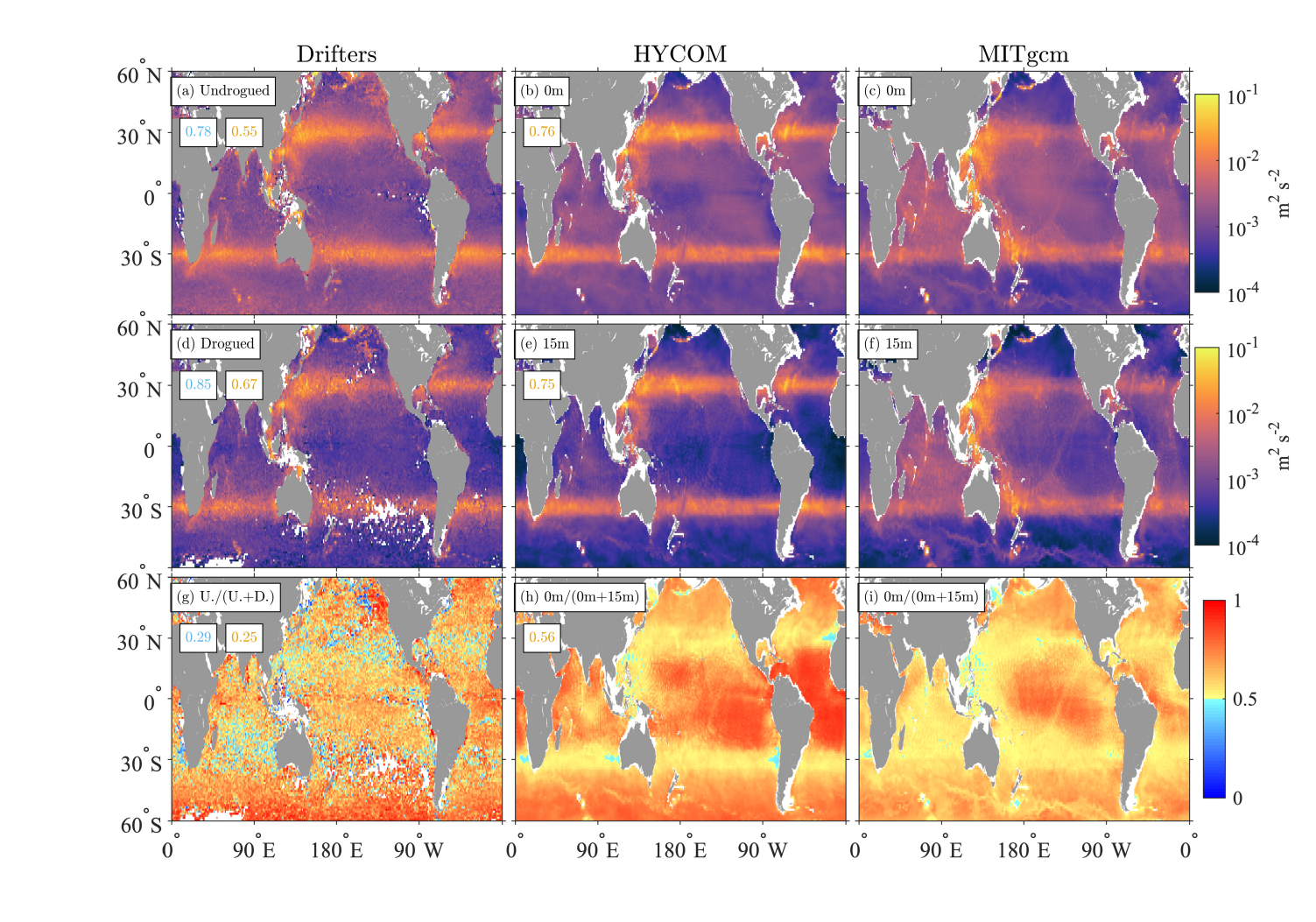}
\caption{Global maps of diurnal ($\pm [0.9,1.1]$ cpd) KE from undrogued and drogued drifters (panels a and d), from HYCOM at 0 m and 15 m (panels b and e), and from MITgcm LLC4320 at 0 m and 15 m (panels c and f). The ratios of undrogued KE/(undrogued KE + drogued KE) for the drifters, and 0 m KE/(0 m KE + 15 m KE) for the HYCOM and MITgcm LLC4320 simulations, are shown in panels g, h, and i, respectively.  The spatial correlations between the drifter maps and HYCOM and MITgcm LLC4320 maps are indicated in the upper left corners of panels a, d, and g, with HYCOM values in cyan to the left of MITgcm LLC4320 values in gold. The spatial correlations between HYCOM and MITgcm LLC4320 maps are indicated in the upper left corners of panels b, e, and h.} 
\label{fig:diurnal_maps}
\end{figure}

The most energetic common spatial feature of the diurnal KE maps for the drifters, HYCOM, and MITgcm LLC4320 at both depth levels (Figure~\ref{fig:diurnal_maps}, panels a to f) corresponds to wind-driven near-inertial motions around $\pm 30^{\circ}$ latitude.  MITgcm LLC4320 therefore underestimates diurnal motions around these latitudes where diurnal and near-inertial motions overlap.  Another common spatial feature is a global pattern associated with baroclinic tidal energy constrained equatorward of $\sim30^{\circ}$.  The latter pattern is clearly seen in the models, but is less visible in the drifters because of the higher noise level in drifter data.  MITgcm LLC4320 appears to overestimate the tidally-forced diurnal motions.  Some faint but distinguishable diurnal KE features exist at both depth levels along the Agulhas Return Current and the Antarctic Circumpolar Current, in both models.  These features are not as visible in the drifter maps, perhaps because of noise levels and poor sampling.  The spatial correlations at both levels between models (0.76 and 0.75) indicate that the models capture similar KE patterns.  At both depth levels, the spatial correlations between the drifter results and the model results suggest that HYCOM better captures the observations (0.78 and 0.85) than MITgcm LLC4320 does (0.55 and 0.67).

The maps of diurnal band vertical structure proxy ratio for the models (Figure~\ref{fig:diurnal_maps}, panels h and i) are generally similar, and exhibit a modest correlation between each other (0.56).  The models display some modest correlation with the drifter ratio map (0.29 for HYCOM and 0.25 for MITgcm LLC4320).  Similar patterns of relatively high proxy ratio values are visible in the equatorial regions and Southern Ocean, for both the drifter and model maps.  The vertical structure proxy ratio is relatively low around $\pm 30^\circ$.  In this frequency and latitude band, KE is dominated by wind-driven near-inertial motions, which as discussed in the previous section have a measurable but small amount of vertical structure.

\begin{figure}
\centering
\noindent\includegraphics[width=\textwidth]{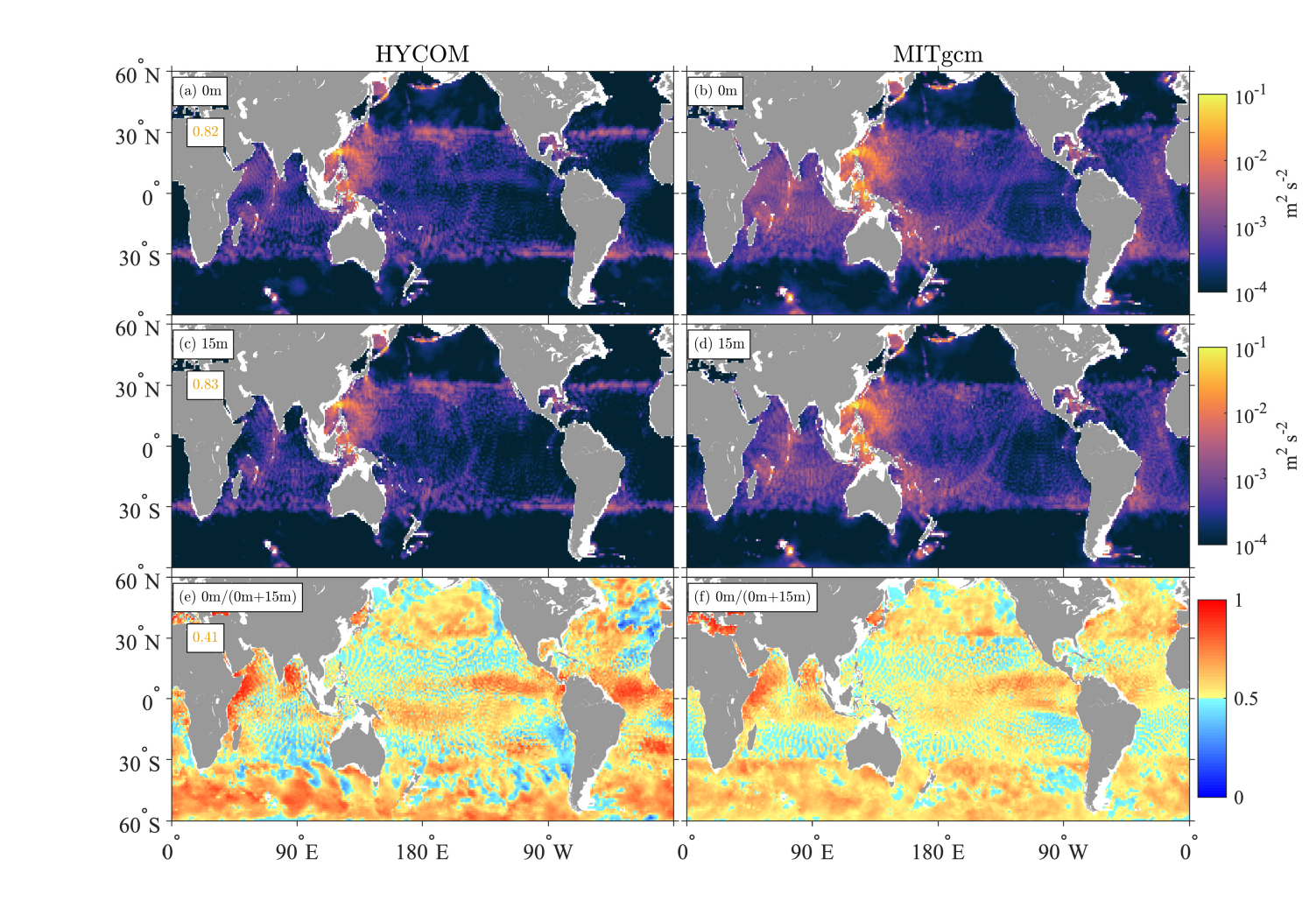}
\caption{Global maps of diurnal KE calculated from harmonic analysis of diurnal tidal constituents K$_{1}$ and O$_1$ for HYCOM at 0 m and 15 m (panels a and c), and from MITgcm LLC4320 at 0 m and 15 m (panels b and d). The ratios of 0 m KE/(0 m KE + 15 m KE) for the HYCOM and MITgcm LLC4320 simulations are shown in panels e and f, respectively.  The spatial correlations between HYCOM and MITgcm LLC4320 maps are indicated in the upper left corners of panels a, c, and e.}
\label{fig:diurnal_harmonicanalysis_maps}
\end{figure}

In the model results, we can separate the stationary (sometimes referred to as phase-locked or coherent) diurnal tidal motions from other diurnal motions using a tidal harmonic analysis, here of the two largest constituents (K$_{1}$ and O$_1$).  Diurnal KE maps obtained from harmonic analysis, shown in Figure~\ref{fig:diurnal_harmonicanalysis_maps} (panels a to d), display enhanced regions near 30$^{\circ}$, but these do not dominate the maps as dramatically as in Figure~\ref{fig:diurnal_maps} because the wind-driven KE, which is not stationary, is not captured in the harmonic analysis.  Substantial diurnal tidal KE regions are also noticeable equatorward of 30$^{\circ}$, especially in the well-known diurnal tide hotspot in the western Pacific or along beams of baroclinic motions emanating from topographic features \cite{Wang2021}; this location is also prominent in drifter data, e.g, Figures \ref{fig:diurnal_maps}a and d.  Except in specific locations such as the Sea of Okhotsk, Aleutian Island chain, the Campbell Plateau, and Kerguelen Plateau, diurnal KE computed from tidal harmonic analysis drops off steeply poleward of 30$^{\circ}$, the cutoff latitude separating freely propagating from evanescent diurnal tidal motions.  For both models, some diurnal KE is captured poleward of $30^\circ$ by the harmonic analysis around some prominent topographic features in the Southern Ocean (for instance around the Kerguelen Plateau and the Campbell Plateau) or within the Alaskan Archipelago in the North Pacific.  These features correspond to strong barotropic tidal currents, and some evanescent diurnal baroclinic tidal energy, unable to propagate far from their generation regions.  The features seen in the total diurnal KE maps (e.g., Figures \ref{fig:diurnal_maps}e-f) within the Agulhas Return Current and the Antarctic Circumpolar Current are absent in these harmonic analysis maps, suggesting that these are wind-driven in nature.

The vertical structure proxy ratio maps for the diurnal KE from the harmonic analysis (Figure ~\ref{fig:diurnal_harmonicanalysis_maps}, panels e and f) are complex.  The ratios appear to be the result of superimposed long-wavelength patterns of barotropic tidal motions and short-wavelength patterns of baroclinic tidal motions within the tropical regions, and superimposed patterns of barotropic tidal motions and residual wind-driven motions at extratropical latitudes.   
The spatial correlations of the model diurnal KE maps (Figures \ref{fig:diurnal_harmonicanalysis_maps}a and c) from the harmonic analysis (0.82 for 0 m and 0.83 for 15 m) are slightly higher than for the total diurnal KE maps (0.76 and 0.75, for 0 and 15 m, respectively; Figures \ref{fig:diurnal_maps}b and e). In contrast, the spatial correlation between models of the diurnal tidal ratio maps is lower (0.41, Figure~\ref{fig:diurnal_harmonicanalysis_maps}e) than for the total diurnal KE ratio maps (0.56, Figure~\ref{fig:diurnal_maps}h).  

\begin{figure}
\centering
\noindent\includegraphics[width=\textwidth]{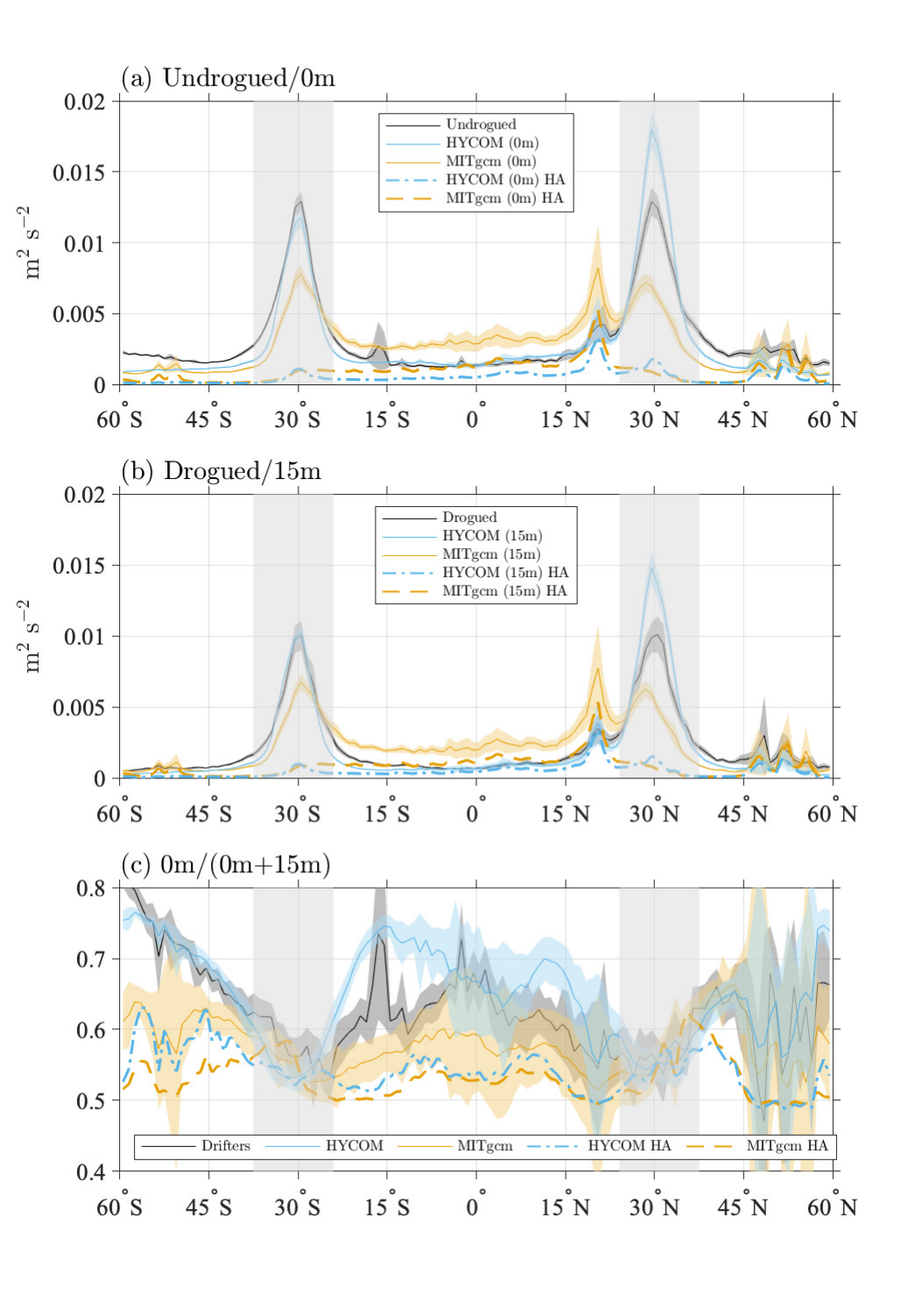}
\caption{Zonally averaged diurnal ($\pm [0.9,1.1]$ cpd) KE from (a) undrogued drifters and 0~m model levels, and (b) drogued drifters and 15~m model levels.  Zonal averages of undrogued KE/(undrogued KE + drogued KE) for the drifters, and 0 m KE/(0 m KE + 15 m KE) in HYCOM and MITgcm LLC4320, are shown in (c).  The zonally averaged diurnal KE from harmonic analysis ("HA", in legend) is also plotted for HYCOM (dash-dotted curve) and MITgcm LLC4320 (dashed curve). The gray shaded region in each panel indicates the latitudes where the diurnal and near-inertial frequency bands overlap (see Fig.~\ref{fig:frequency_latitude_spectra}). The shading around each curve corresponds to two standard errors of the calculated 1$^\circ$ zonal averages from the geographical maps.}
\label{fig:diurnal_zavg}
\end{figure}

Zonal averages (Figure~\ref{fig:diurnal_zavg}) confirm that HYCOM lies closer to the diurnal peaks in the drifter results near 30$^{\circ}$S and 30$^{\circ}$N, while MITgcm LLC4320 diurnal energy is too weak in these peak regions.  Equatorward of these peaks, MITgcm LLC4320 diurnal energy is generally too strong relative to drifter results, especially near 20$^{\circ}$N, the latitude of the northwestern Pacific internal tide hotspot, while HYCOM KE is generally comparable to drifter values.  

The zonally averaged diurnal band vertical structure proxy ratios in the models (Figure~\ref{fig:diurnal_zavg}c) are complicated, as anticipated, and are revealing of model strengths and weaknesses.  The drifter ratio is relatively high near the equator and at high latitudes.  The MITgcm LLC4320 diurnal KE ratio follows the drifter ratio comparatively well over low- and mid-latitudes but is too low at high latitudes.  The HYCOM diurnal KE ratio follows higher latitude drifter values more closely, and also follows the mid-latitude drifter ratio well, but is noticably higher than the drifter ratio over many latitudes equatorward of about 20$^{\circ}$.  Yet, recall that HYCOM tracks the absolute drifter KE values equatorward of 20$^{\circ}$ more closely than MITgcm LLC4320 does (panels a and b).  The zonally averaged vertical structure proxy ratio for the diurnal tidal harmonic analysis KE lies relatively close to 0.5 over all latitudes.  The relatively low ratios, and relatively low KE in the harmonically analyzed diurnal motions, suggests that stationary diurnal tidal motions are not a major contributor to the vertical structure of KE seen in the diurnal band.  We will connect this to an analysis of vertical modes in the discussion section.

\subsection{Semidiurnal KE}

\begin{figure}
\centering
\noindent\includegraphics[width=\textwidth]{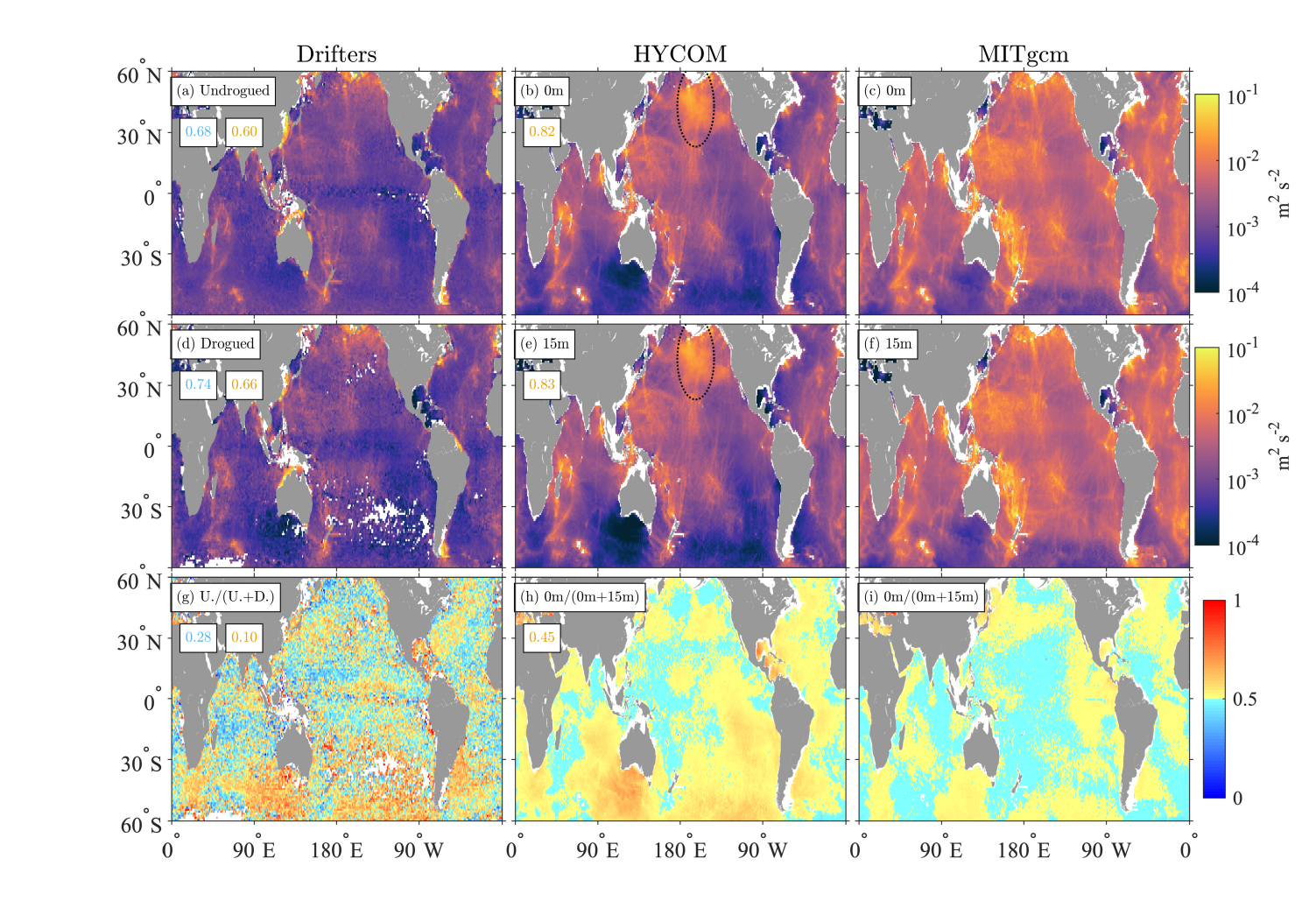}
\caption{Global maps of semidiurnal ($\pm [1.9,2.1]$ cpd) KE from undrogued and drogued drifters (panels a and d), from HYCOM at 0 m and 15 m (panels b and e), and from MITgcm LLC4320 at 0 m and 15 m (panels c and f). The ratios of undrogued KE/(undrogued KE + drogued KE) for the drifters, and 0 m KE/(0 m KE + 15 m KE) for the HYCOM and MITgcm LLC4320 simulations, are shown in panels g, h, and i, respectively.  The spatial correlations between the drifter maps and HYCOM and MITgcm LLC4320 maps are indicated in the upper left corners of panels a, d, and g, with HYCOM values in cyan to the left of MITgcm LLC4320 values in gold. The spatial correlations between HYCOM and MITgcm LLC4320 maps are indicated in the upper left corners of panels b, e, and h.  The North Pacific region of numerical instability in HYCOM \cite{Buijsman2016} is indicated in panels b and e.}  
\label{fig:semidiurnal_maps}
\end{figure}

Global maps of semidiurnal kinetic energy (Figure~\ref{fig:semidiurnal_maps}, panels a to f) display known hotspots of semidiurnal internal tide motions near, for instance, Hawai'i, the French Polynesian islands, and the western Pacific.  The KE values are substantially higher in the simulations, especially MITgcm LLC4320, than in the drifter observations.  The hotspots are visible in the drifter maps, but less so because of the higher noise level in the drifter data.  As highlighted in panels b and e, in HYCOM, the semidiurnal kinetic energy is spuriously large in a patch of the high-latitude North Pacific, south of the Aleutians, due to a known numerical instability \cite{Buijsman2016}.  As is generally seen in other frequency bands, spatial correlations between model KE maps at both 0 and 15 m are higher (0.82 and 0.83) than the spatial correlations between either model KE map and the drifter KE maps, which range between 0.60 and 0.74.  As in other frequency bands, spatial correlations between HYCOM and drifter semidiurnal KE maps are slightly higher than spatial correlations between MITgcm LLC4320 and drifter semidiurnal KE maps.

\begin{figure}
\centering
\noindent\includegraphics[width=\textwidth]{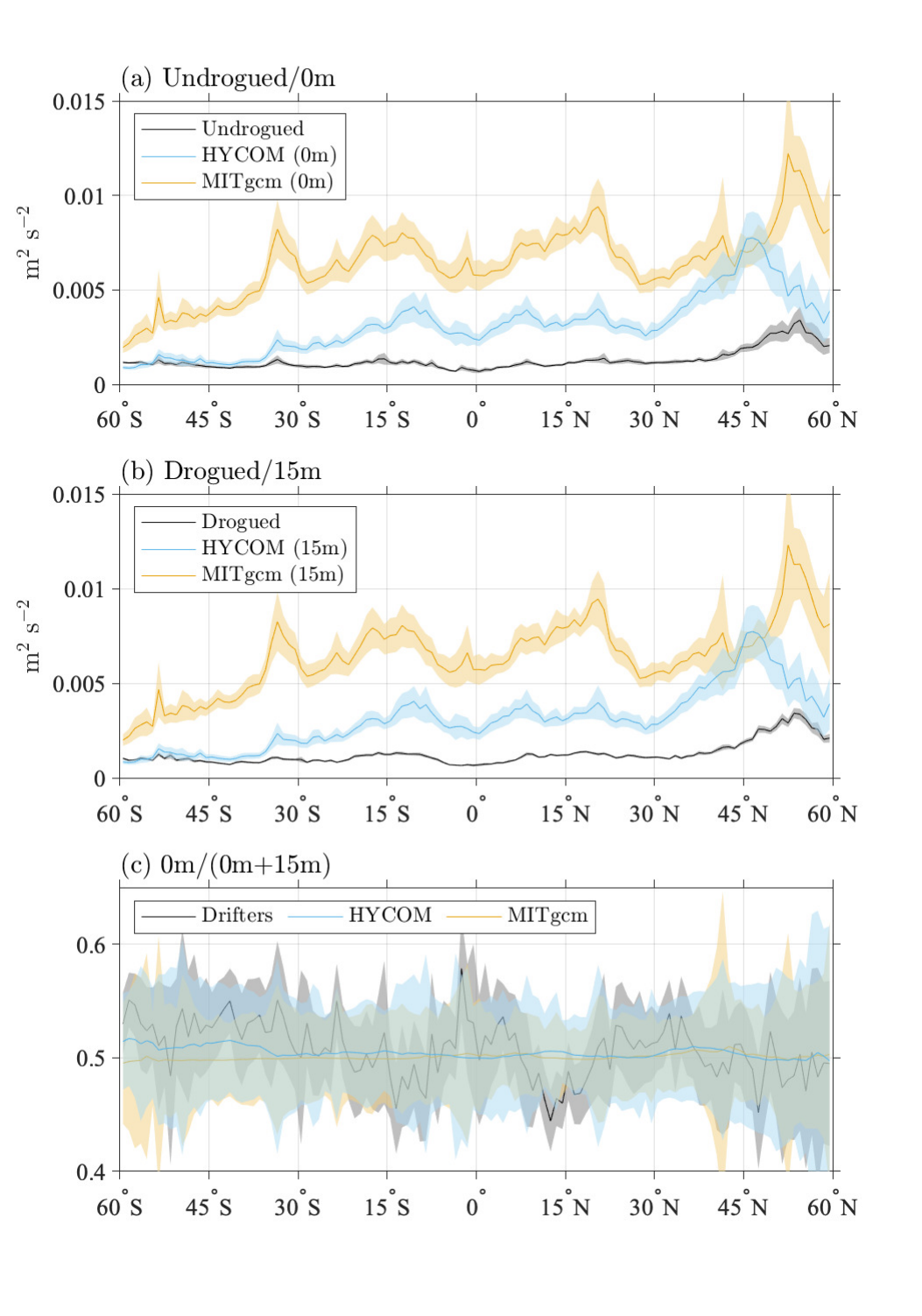}
\caption{Zonally averaged semidiurnal ($\pm [1.9,2.1]$ cpd) KE from (a) undrogued drifters and 0~m model levels, and (b) drogued drifters and 15~m model levels.  Zonal averages of undrogued KE/(undrogued KE + drogued KE) for the drifters, and 0 m KE/(0 m KE + 15 m KE) for HYCOM and MITgcm LLC4320, are shown in (c).  The shading around each curve corresponds to two standard errors of the calculated 1$^\circ$ zonal averages from the geographical maps.}
\label{fig:semidiurnal_zavg}
\end{figure}

The vertical structure proxy ratio maps for the semidiurnal KE band differ substantially from ratio maps in other frequency bands. For both models (panels h and i), the ratio values are generally close to 0.5 indicating low differences of KE levels between the surface and 15 m.  However, the spatial patterns are rather different, resulting in a modest correlation between the two model ratio maps (0.45). The ratio map for the drifters is again noisy (panel g), yet shows, as in the model maps, values closer to 0.5 than in other frequency bands.  The spatial correlation between the drifter semidiurnal ratio map and the HYCOM semidiurnal ratio map is substantially larger than for MITgcm LLC4320 (0.28 compared to 0.10).  This marked difference may be due to the ability of HYCOM to represent higher ratio values in the Southern Ocean and Gulf of Mexico, as is seen in the drifter observations. 
In the zonal averages (Figure~\ref{fig:semidiurnal_zavg}, panels a and b), as demonstrated by \citeA{Yu2019}, MITgcm LLC4320 KE is higher than drifter KE over all latitudes, by a factor of up to four.  Over most latitudes, HYCOM lies closer to the drifter KE than MITgcm LLC4320 does.  A notable exception to this pattern is seen in northern hemisphere high latitudes, where the numerical instability in North Pacific HYCOM, mentioned earlier, is exhibited.  The semidiurnal KE in HYCOM is still substantially larger than the drifter KE, in contradistinction to the closer agreement seen in comparisons of HYCOM internal tide SSH signatures with altimetry \cite{Ansong2015,Buijsman2020}.  This illustrates the value of comparing models with velocity observations as well as SSH observations.  

In contrast with results in other frequency bands, the vertical structure proxy ratios in the semidiurnal band are consistent with values of 0.5, to within error bars (Figure~\ref{fig:semidiurnal_zavg}c), implying that there is on average no vertical structure in the semidiurnal band compared to the other bands.  However, the large error bars on the drifter and model semidiurnal results may encompass the possibility of a small spatially-varying shear.

\begin{figure}
\centering
\noindent\includegraphics[width=\textwidth]{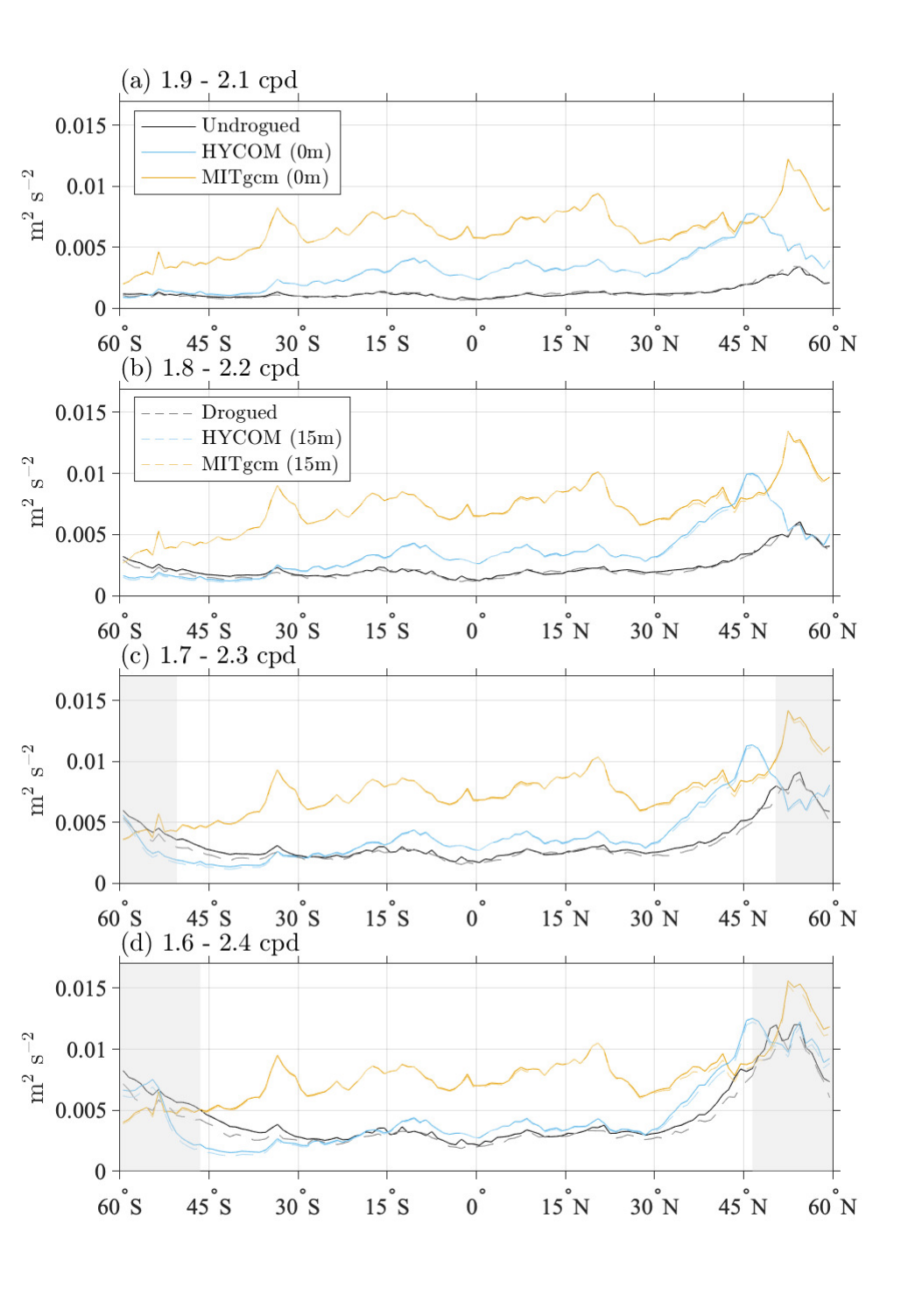}
\caption{Zonally averaged semidiurnal KE estimates from drifters and model levels calculated by gradually widening the band of frequency integration from $\pm[1.9,2.1]$ cpd to $\pm[1.6, 2.4]$ cpd as indicated in the title of each subplot. Uncertainty estimates are omitted here for clarity. In panels c and d, the gray shaded areas indicate the latitudes where the frequency bands of integration overlap with the near-inertial bands. The legend boxes in panels a and b are applicable to all panels.}
\label{fig:semi_wide}
\end{figure}

Motivated by the ``wider and lower" semidiurnal tidal peaks in drifter spectra compared with model spectra (see insets of Figure \ref{fig:globalavg_spectra}), we examine in Figure~\ref{fig:semi_wide} how the model vs. drifter semidiurnal zonal average comparisons change if the $\pm [1.9,2.1]$ cpd semidiurnal band defined by \citeA{Yu2019} is widened to $\pm [1.6, 2.4]$ cpd in incremental $0.1$ cpd steps.  As the definition of the semidiurnal band is widened, the drifter KE levels rise gradually, such that, over many latitudes, the agreement between HYCOM and drifters becomes fairly close for the widest definitions of the band (Figure~\ref{fig:semi_wide}, panels c and d).  Again, the northern hemisphere high latitudes, where the numerical instability in North Pacific HYCOM is present, represent an exception to this pattern.  Over most latitudes, MITgcm LLC4320 KE levels sit well above the drifters--typically by a factor of about two--for all definitions of semidiurnal band employed in Figure~\ref{fig:semi_wide}.  By taking the wider nature of Lagrangian drifter spectra into account, these results suggest that the tidal currents in HYCOM may actually lie closer to the drifter values than the results in Figure~\ref{fig:semidiurnal_zavg} indicated.   

\section{Discussion}

The difference in HYCOM and MITgcm LLC4320 near-inertial KE levels likely arises from the higher-frequency wind forcing in HYCOM (3 hours) vs. MITgcm LLC4320 (6 hours).  \citeA{Rimac2013} demonstrated that near-inertial KE in models is highly sensitive to the atmospheric forcing update intervals, with hourly forcing yielding KE three times higher than 6-hourly forcing.  \citeA{Flexas2019} found relatively low wind-power input to the near-inertial motions in MITgcm LLC4320, due to the 6-hourly updates in the wind forcing.  The HYCOM near-inertial KE, though higher than MITgcm LLC4320, is still too weak in the Pacific Ocean relative to drifter observations.  Aside from insufficiently frequent atmospheric forcing updates, another potential cause of the model-drifter discrepancy in the near-inertial band is that the model outputs are of one year duration, while the drifter data span multiple years. 

The largest discrepancy between models and drifter observations is in low-frequency flows at the equator; note the different y-axis scales in the low-frequency zonal average plot vs. the zonal average plots in other frequency bands.  It is intriguing that HYCOM, with its more frequent atmospheric forcing, lies closer to the drifters in the equator than MITgcm LLC4320 does.  The Intertropical Convergence Zone, or ITCZ, features rain and wind variability at high frequencies (sub-daily time scales) and small horizonal scales \cite{Clayson2019,Thompson2019}.  This high-frequency wind and rain variability could potentially impact vertical momentum transfer into the ocean in the near-equatorial region.  Thus, it is possible that atmospheric update intervals would make a difference to equatorial low-frequency dynamics as well as mid-latitude near-inertial dynamics.  Fully probing the equatorial dynamics at work here is a subject for future research.

MITgcm LLC4320 has stronger diurnal tidal motions than HYCOM, probably because of the lack of parameterized topographic wave drag in MITgcm LLC4320. The wave drag in HYCOM is tuned for semidiurnal tides, not diurnal tides, which may as a result be over-damped in that simulation.

The model diurnal tide harmonic analysis KE values are significantly weaker than the model total diurnal KE values computed from integration of the frequency spectra in the diurnal band.  This confirms that the diurnal band consists of motions other than stationary internal tides (nonstationary diurnal tides, diurnal cycling of Ekman and submesoscale flows, and, in latitudes near 30$^{\circ}$, near-inertial flow).

As indicated in Section 2.2, MITgcm LLC4320 had an overly large tidal forcing, and lacked the Self-Attraction and Loading (SAL) term, but these effects are not large enough to explain the discrepancy between MITgcm LLC4320 and drifter observations in the semidiurnal band.  In future work, more direct comparisons of HYCOM and drifters, in which HYCOM is ``seeded" with synthetic drifters, are warranted to investigate more comprehensively the  biases introduced by the drifter Lagrangian sampling \cite{ZaronElipot2021}.  If the closer agreement between HYCOM and drifters in the semidiurnal band holds up under further scrutiny, it would suggest, as in \citeA{Ansong2015}'s analysis of internal tide SSH signals, that parameterized topographic wave drag (or some other plausible damping mechanism) is necessary for achieving agreement between semidiurnal tidal KE in models and observations.

The general agreement of the vertical structure proxy ratio in models and drifters, seen in spatial maps and zonal averages, supplies some confidence that the vertical structure proxy ratio computed from drifters is meaningful, noisy though it may be.  Mode values of these ratios, for drifter observations and both models, over all frequency bands displayed in our 1$^\circ$  by 1$^\circ$ global maps, are given in Table~\ref{tab:vertical_structure}.  The global mode values are consistent with the presence of vertical structure in the low-frequency and diurnal bands, and, to a lesser extent, the near-inertial band.  The semidiurnal band has less vertical structure, with most reported values being near 0.5.       

\begin{table}[]
\begin{tabular}{l|l|l|l|l}
         & Low-frequency & Near-inertial & Diurnal & Semidiurnal \\\hline
Drifters &  0.61             &         0.55             &         0.65          &            0.53  \\\hline
HYCOM    & 0.57          &            0.52         &             0.68       &                0.50  \\\hline
MITgcm   &  0.54          &            0.54          &            0.55        &               0.50 \\\hline
\end{tabular}
\caption{Mode values of the ratio of undrogued KE/(undrogued KE + drogued KE) for the drifters, and 0 m KE/(0 m KE + 15 m KE) for HYCOM and MITgcm LLC4320, in four different frequency bands. A value of 0.5 indicates that the KE is the same at the two depth levels.}
\label{tab:vertical_structure}
\end{table}

To aid in interpreting the vertical structure of motions that are not directly wind-driven, such as diurnal tides, semidiurnal tides, and the low-frequency eddying general circulation (the latter arising indirectly from wind forcing), it is useful to compute the first three vertical modes \cite<e.g.,>[among many]{Gill1982,Arbic2018}, for semidiurnal, diurnal, and zero frequency (the latter representing the quasi-geostrophic limit).  We use five different deep-ocean profiles in a US Navy climatology \cite{Helber2013}, at 10$^{\circ}$S, 206$^{\circ}$E and at four different latitudes (50$^{\circ}$N, 10$^{\circ}$N, 6$^{\circ}$N, and 50$^{\circ}$S) along 210$^{\circ}$E.  The values of the vertical structure proxy ratio computed from these low vertical modes are greater than 0.5, but only by small amounts (ranging from 0.5000-0.5012), less than the near-inertial values seen in Figure~\ref{fig:near_inertial_zavg}c and much less than the low-frequency and total diurnal values seen respectively in Figures~\ref{fig:lowf_zavg}c and~\ref{fig:diurnal_zavg}c.  The low vertical structure proxy ratios in the vertical normal mode analysis are consistent with the lack of vertical structure seen in our semidiurnal results and diurnal tidal harmonic analysis results, both of which are almost entirely due to tidal motions.  Semidiurnal and especially diurnal tidal currents do have a surface-intensified profile \cite[their Figure 5]{Timko2013}, but this surface intensification takes place over vertical scales that are significantly larger than 15 m.  

The lack of vertical structure between 0 and 15 m in the low vertical mode analysis is inconsistent with the vertical structure proxy values exceeding 0.5 in the low-frequency and diurnal band results.  Additional physics, such as wind-driven Ekman and other motions, is taking place in the diurnal and low-frequency bands (and in the near-inertial band).  As discussed in Section 2.3, motions that are directly wind-driven are expected to exhibit some vertical structure in the upper ocean.  The proxy ratio values greater than 0.5 in the diurnal band results (Figure~\ref{fig:diurnal_zavg}c) are likely due to non-tidal effects, such as near-inertial flows (in latitudes where they overlap), diurnal cycling in Ekman flows \cite{Price1986,Price1999,SunSun2020}, and diurnal cycling in submesoscale flows \cite{Sunetal2020}.  Regarding the latter possibility, however, we note that the grid spacings in the global models examined here are not sufficient to fully resolve submesoscale eddies \cite{Capet2008}.  

The ``interior quasi-geostrophic'' component \cite{Lapeyre2006} of low-frequency large-scale currents and mesoscale eddies is dominated by barotropic and low baroclinic modes \cite{Wunsch1997}, or, in an alternative view, by a ``surface mode'' that is strongest at the surface and approaches zero flow at the seafloor \cite{Lacasce2017}, due to the influence of bottom topography \cite{Lacasce2017} and/or bottom and topographic wave drag \cite{Arbic2004,Trossman2017}.  The surface mode described by \citeA{Lacasce2017} decays with depth at the ocean surface, as do the traditional low baroclinic modes.  In addition, low-frequency motions may also have a substantial ``surface quasi-geostrophic'' component \cite{Lapeyre2006,LaCasce2015}, which is surface intensified and may therefore contribute to the weaker motions at 15 m depth relative to the surface.  However, the simplest explanation for the proxy ratio values greater than 0.5, seen in the low-frequency band (Figure \ref{fig:lowf_zavg}c), is that they are due to Ekman flows, which exhibit substantial variation over short vertical scales \cite{Elipot2009,LillyandElipot2021}. 

\section{Summary and conclusion}

Near-surface ocean kinetic energy (KE) is an important factor in a variety of problems, including but not limited to air-sea interaction, pollution transport, and satellite mission planning.  Such applications require better quantification and understanding of the space-time variability of near-surface oceanic KE, including its frequency dependence and vertical structure \cite{ElipotWenegrat2021}. 

We have compared the KE in two widely used global high-resolution general circulation simulations (HYCOM and MITgcm LLC4320) to drifter observations.  We compare the sea surface (0 m) KE in the models to KE from undrogued drifters, and the KE at 15 m depth in the models to KE from undrogued drifters.  We also compare a vertical structure proxy ratio, computed from 0 m KE/(0 m KE + 15 m KE) in the models and from undrogued KE/(undrogued KE + drogued KE) in the drifter observations.  We compare the KE and the vertical structure proxy ratio across a wide range of frequencies, enabled by new drifter technologies and analyses that allow for high-frequency mapping, and by the inclusion of astronomical tidal forcing and relatively high atmospheric forcing updates in HYCOM and MITgcm, that activates tidal and near-inertial flows alongside lower-frequency flows such as Ekman flows and the eddying general circulation.  

Our goal is a descriptive paper focused on global maps of KE and the vertical structure proxy ratio.  The uniqueness and computational cost of the simulations used here prevent us from conducting sensitivity analyses for the comparison to observations.  The drifter dataset as well is unique and quite different in nature from the models, meaning that some of the comparisons made here are not as ``apples-to-apples" as one might have preferred.  Some inconsistencies between analyses of the models and drifters, and examples of important contributing processes that should be examined in more detail, are listed below: 
\begin{itemize}
    \item Generating robust statistics and geographical coverage from drifter data requires us to use many years of observations.  We use more than 30 years here.
    \item At the same time, the computational expense of the simulations, and the fact that they are run at different centers with no coordination between them, requires us to use a small number of years, and makes it difficult to use the same year from each model simulation.  We use one year of output from the HYCOM and MITgcm LLC4320 simulations--with start dates of 1 January 2014 and 12 November 2011, respectively.
    \item The KPP parameters, which exert strong control on vertical structure, differ between the two simulations.
    \item The vertical grid resolution also differs between HYCOM and MITgcm LLC4320.
    \item The atmospheric products used to force the two simulations are different, and their update frequency is different (3-hourly NAVGEM for HYCOM, 6-hourly ECMWF for MITgcm LLC4320).
    \item The model results are Eulerian, whereas the drifter results are Lagrangian.
    \item The frequency dependence of the noise in the drifter observations, in particular the undrogued drifters, is not yet characterized.
    \item We do not examine the important topic of seasonality.
    \item The MITgcm LLC4320 simulation had a few mistakes in the tidal forcing, and did not employ a parameterized topographic wave drag as HYCOM did.
\end{itemize}

Our study follows that of \citeA{Yu2019}, who used only MITgcm LLC4320 but whose study otherwise likewise suffered from all of the problems noted above.  The three-way comparison between drifters and two models allows us to assess the strengths and weaknesses of all three products.  Following \citeA{Yu2019}, who compared KE in the MITgcm LLC4320 simulation to drifters, we find that near-inertial motions in MITgcm LLC4320 are too weak while semidiurnal tidal motions are too strong.  Here we find that the HYCOM KE values lie closer to the drifters in both the near-inertial and semidiurnal tidal bands, but for different reasons.  In the near-inertial band, HYCOM is stronger than MITgcm LLC4320 due to more frequently updated atmospheric forcing fields.  In the semidiurnal tidal band, HYCOM is weaker than MITgcm LLC4320, due primarily to the parameterized topographic internal wave drag, which simulates the energy lost due to unresolved wave generation and breaking processes, and which is employed in HYCOM but not in MITgcm LLC4320.  While HYCOM semidiurnal tidal KE lies closer to the drifters than MITgcm LLC4320 KE does, the HYCOM semidiurnal KE is still stronger than the drifter KE if the \citeA{Yu2019} definition of the semidiurnal band is employed.  However, if we widen the definition of semidiurnal band from that employed in \citeA{Yu2019}, the HYCOM tidal KE lies closer to the drifters over most latitudes, while MITgcm LLC4320 semidiurnal KE is too high.  Widening the definition of the semidiurnal band to accommodate comparisons to drifters may be justified, due to the inherently ``wider" nature of Lagrangian spectra relative to Eulerian spectra \cite{ZaronElipot2021}.  To be more sure of this interpretation, both HYCOM and MITgcm LLC4320 could be seeded with numerical particles and the resulting Lagrangian velocity spectra could be more directly compared to drifter spectra from the actual ocean.  This computationally expensive undertaking is left as a topic for future investigation. 

Our conclusion that damping parameterizations are necessary for attaining realistic semidiurnal tidal kinetic energy levels is consistent with the results of \citeA{Ansong2015}.  The latter paper demonstrated that the sea surface height signature of internal tides in HYCOM is closer to altimetry observations when the HYCOM simulations contain a wave drag than when they do not.  The work here, comparing HYCOM and MITgcm LLC4320 simulations with drifter observations, suggests that the lack of parameterized wave drag in LLC4320 leads to overly large tidal kinetic energies.  Indeed, the contrast between simulations with and without wave drag is greater for tidal kinetic energy than it is for internal tide sea surface height.  Some caution regarding this interpretation is warranted, however, as we are comparing two different models.

The diurnal band is complicated because of overlapping processes.  Diurnal motions include diurnal tides, diurnal cycling of Ekman and submesoscale flows, and, in latitudes where the frequency-band definitions overlap ($\sim$ 30$^{\circ}$), near-inertial flows.  The stationary component of diurnal tides can be computed separately from other diurnal motions via a tidal harmonic analysis.  Maps and zonal averages indicate that stationary diurnal tides are a relatively small component of the KE seen in the diurnal band. 
    
In additional results, we have shown that HYCOM lies closer to drifters in the diurnal band (with the notable exception of the vertical structure proxy ratio at low latitudes), that low-frequency ($<$0.5 cpd) motions in both models are generally too weak, especially near the equator, that both models suffer from weak near-inertial motions in the northern mid-to-high latitudes, and that numerical instabilities yield overly large tidal HYCOM semidiurnal KE in the North Pacific.  

HYCOM has higher spatial correlations with drifter KE than MITgcm LLC4320 does, across a wide variety of frequency bands.  This result is consistent with \citeA{Luecke2020}'s finding that the spatial correlations between HYCOM and mooring observations of KE and temperature variance are higher than the correlations between MITgcm LLC4320 and mooring observations.

With some exceptions, in the maps and zonal averages, the models capture the latitude- and frequency- dependence of the vertical structure proxy ratio relatively well.  The low-frequency, near-inertial, and diurnal bands display significant vertical structure, while the semidiurnal band shows little vertical structure.  Ekman flows, and their diurnal cycling, likely explain some of the vertical structure seen in the low-frequency and diurnal bands.

The broad approach employed here has left us with several unsolved problems and questions.  For instance, the near-surface KE in the diurnal band, which is forced by the astronomical tidal potential and the diurnal solar heating cycle, requires much more examination than we provided here.  The large differences between low-frequency equatorial flows in the models and drifter observations represent a first-order problem for future work.  In addition, the vertical structure of the upper ocean velocities throughout the entire mixed layer (not just at 0 and 15 m), and the sensitivity of these velocities to the vertical grid spacing and KPP parameters, is another topic that deserves more attention.  The geographical and frequency dependence of upper ocean flows and their vertical structure is indeed a rich topic for research.    

\appendix
\section{Drifter velocity filtering}

To filter the drifter velocity time series and eventually obtain KE maps, we consider all trajectories of the drifter hourly dataset (version 1.04c) that are 4-day long and longer. We have chosen a time series filter which is a continuous wavelet transform using a generalized Morse wavelet \cite{lilly2017element}. For the diurnal and semidiurnal cases, to achieve a bandpass filter, the frequency spectrum of the generalized Morse wavelet is chosen to be centered on the middle frequency of the desired frequency band (1 and 2 cpd for the diurnal band and semidiurnal bands, respectively), and the half-power bandwidth of the wavelet spectrum is set to approximately match the desired interval for the bandpass ([0.9, 1.1] cpd and [1.9, 2.1] cpd for the diurnal and semidiurnal bands, respectively). Following the notation of \citeA{lilly2012generalized}, such characteristics of the wavelet are achieved by setting the Morse wavelet parameter $\gamma$ to 3, and the parameter $\beta$ to $(\nu/\Delta \nu)^2/\gamma$ where $\nu$ is the center frequency of the frequency-domain wavelet in each case, while $\Delta \nu$ is its half-power bandwidth (0.1 cpd in our case). Such calculation of the wavelet parameters however relies on a quadratic approximation of the frequency-domain wavelet to determine the half-power point frequencies. Practically, this wavelet transform and filtering method is implemented with the function \texttt{wavetrans} of the Matlab toolbox jLab \cite{Lilly2021}. An online course describes this specific method of using the wavelet transform as a narrow bandpass filter \cite{jonathan_m_lilly_2022_5977995}.

For the low-frequency band, in contrast to the approach used for the models, we explicitly lowpass the drifter velocities before calculating the variance in spatial bins. This lowpass variance is added to the mean velocity field squared to obtain the low-frequency KE. The lowpass filter is also achieved here by using a wavelet transform, but by setting the Morse wavelet parameters $\gamma$ to 3 and $\beta$ to 0, and by setting the half-power point frequency to 0.5 cpd \cite[section IV. A.]{lilly2009higher}. 

For the near-inertial band, we also apply a wavelet transform method, but one that continuously changes the center frequency of the wavelet to follow the local inertial frequency as the drifter latitude changes. This specific method is now coded in the function \texttt{inertialextract}, also part of the Matlab toolbox jLab \cite{Lilly2021}. The function is used by specifying the wavelet duration parameter $P = \nu/\Delta \nu$. With $\nu= -f$ and $\Delta \nu = -0.1 f$, we obtain $P=10$. The function is implemented by passing as arguments the times, longitudes, and latitudes of a drifter trajectory. The function calculates the local inertial frequency based on the latitude, and subsequently returns the instantaneous complex-valued inertial oscillation amplitudes in kilometers. These displacement amplitude time series are subsequently differentiated with respect to time to estimate the corresponding near-inertial velocities, using a central difference scheme. Because the wavelet transform operation implemented in \texttt{inertialextract} is analytic, it filters the energy only on one side of the frequency domain, whichever one corresponds to the anticyclonic side, depending on the sign of the latitude. Because we are seeking to extract the KE not only around the inertial, anticyclonic, frequency ($-f$), but also around its opposite, cyclonic, frequency ($f$), the function \texttt{inertialextract} is applied a second time by negating the longitude time series of each drifter. This amounts to a complex conjugation of the time series, which interchanges the cyclonic and anticyclonic sides of the spectrum. As for the model analyses, the spatial variance of the anticyclonic and cyclonic frequencies are summed to obtain the total near-inertial KE. Conceptually, this method is equivalent to the method known in oceanography as complex demodulation \cite{EmeryT01} but with the use of a filter that is specifically chosen for its property of analyticity; see e.g. \cite{lilly2009higher}, Section IIB for a discussion of the importance of this property. 

\section{Open research}

This paper uses software made available by the Pangeo project, by jLab \cite{Lilly2021} and by the ``cmocean" package \cite{Thyng2016}.  MITgcm LLC4320 output is available at \url{https://data.nas.nasa.gov/ecco/data.php?dir=/eccodata/llc_4320}.  The surface drifter data are available at \url{https://www.aoml.noaa.gov/phod/gdp/hourly_data.php}.
HYCOM output can be accessed via the OSiRIS infrastructure.
Co-authors B.K.A. and J.F.S. can be contacted for details on OSiRIS access.  The Matlab code used to process the MITgcm LLC4320 and HYCOM outputs into the results used in this paper, the results from MITgcm 4320, HYCOM, and processed drifter data, and the code used to produce all of the plots in the paper, are provided in \citeA{Arbicetal2022}; see also \url{https://tinyurl.com/2p9eh5yp} (a permanent DOI \url{https://doi.org/10.7302/PTG7-YW20} is pending).

\acknowledgments

The contributions of J.M.B., D.G., and L.G. to this paper were made while they were undergraduate students.

We thank two anonymous reviewers for useful comments that led to improvements in the presentation of this material.  We thank Roger Samelson for useful comments on vertical structure and the need to quantify model vertical resolution in the upper ocean, Saulo Muller Soares for pointing us towards the Rimac et al. (2013) reference, Bob Helber for providing climatologies for the vertical mode computations described herein, J. Thomas Farrar, Baylor Fox-Kemper, Jeroen Molemaker, and Joe LaCasce for useful discussions on vertical structure and diurnal variations in the mixed layer, Luca Centurioni and Eric D'Asaro for discussions on the equatorial results, Jonathan M. Lilly for his advice on using jLab, Tong (Tony) Lee for useful discussions on our near-equatorial results, and J. Thomas Farrar for discussions on the potential usefulness of the results shown here for S-MODE and satellite velocity-measuring missions. 

B.K.A. gratefully acknowledges funding from US National Science Foundation (NSF) grant OCE-1851164 as well as a Research Experience for Undergraduates Supplement (for J.M.B.) to that grant; from Office of Naval Research (ONR) grant N00014-18-1-2544; and from NASA grants NNX16AH79G, NNX17AH55G, and 80NSSC20K1135.  The latter NASA grant supported the participation of D.G. and L.G.  This research was made possible, in part, by the OSiRIS project at University of Michigan and by computing resources provided by the NASA Advanced Supercomputing (NAS) Division of the Ames Research Center.  S.E. and E.D.Z. gratefully acknowledges funding from US NSF Awards 1851166 and 1850961, respectively.  D.M. carried out research at the Jet Propulsion Laboratory, California Institute of Technology, under contract with NASA, with support from the Physical Oceanography and Modeling, Analysis, and Prediction Programs.  A.L.P. and X.Y. are supported by Agence Nationale de la Recherche (ANR) grant number 17-CE01-0006-01.  M.H.A. was supported by Office of Naval Research grant N00014-18-1-2404.  This NRL contribution NRL/7320/JA--2022/5 has been approved for public release. 

This work is a contribution to the S-MODE project, an EVS-3 Investigation awarded under NASA Research Announcement NNH17ZDA001N-EVS3.


%
%


%
%
%
%
%

\end{document}